%% file: paper.tex
\documentclass[printer]{aa}

\usepackage{graphicx}

\usepackage[varg]{txfonts}
\input{definitions.tex}

\begin{document}

\title{Multi-frequency high spectral resolution observations of HCN toward the circumstellar envelope of \ycvn}
\titlerunning{HCN in the circumstellar envelope of \ycvn}
\authorrunning{J. P. Fonfr\'{\i}a et al.}
\author{J.~P.~Fonfr\'ia\inst{1} \and
E.~J.~Montiel\inst{2} \and
J.~Cernicharo\inst{1} \and
C.~N.~DeWitt\inst{2} \and
M.~J.~Richter\inst{3} \and
J.~H.~Lacy\inst{4} \and
T.~K.~Greathouse\inst{5} \and
M.~Santander-Garc\'ia\inst{6} \and
M.~Ag\'undez\inst{1} \and
S.~Massalkhi\inst{1}}
\institute{
  Grupo de Astrof\'isica Molecular, Instituto de F\'isica Fundamental, IFF-CSIC, C/ Serrano, 123, 28006, Madrid (Spain);~\email{jpablo.fonfria@csic.es} \and
  SOFIA-USRA, NASA Ames Research Center, MS 232-12, Moffett Field, CA 94035 (USA)  \and
  Physics Dept. - UC Davis, One Shields Ave., Davis, CA 95616 (USA) \and
  Astronomy Department, University of Texas, Austin, TX 78712 (USA) \and
  Southwest Research Institute, San Antonio, TX 78228 (USA) \and
  Observatorio Astron\'omico Nacional, OAN-IGN, Alfonso XII, 3, E-28014, Madrid (Spain)
}


\abstract{
  High spectral resolution observations toward the low mass-loss rate C-rich, J-type AGB star \ycvn{} have been carried out at 7.5, 13.1 and 14.0~$\mu$m with SOFIA/EXES and IRTF/TEXES.
  Around 130 HCN and H$^{13}$CN lines of bands $\nu_2$, $2\nu_2$, $2\nu_2-\nu_2$, $3\nu_2-2\nu_2$, $3\nu_2-\nu_2$, and $4\nu_2-2\nu_2$ have been identified involving lower levels with energies up to $\simeq 3900$~K.
  These lines have been complemented with the pure rotational lines $J=1-0$ and $3-2$ of the vibrational states up to $2\nu_2$ acquired with the IRAM 30~m telescope, and with the continuum taken with \textit{ISO}.
We have analyzed the data in detail by means of a ro-vibrational diagram and with a code written to model the absorption and emission of the circumstellar envelope of an AGB star.
The continuum is produced mostly by the star with a small contribution from dust grains comprising warm to hot SiC and cold amorphous carbon.
The HCN abundance distribution seems to be anisotropic close to \ycvn{} and in the outer layers of its envelope.
The ejected gas is accelerated up to the terminal velocity ($\simeq 8$~\kms) from the photosphere to $\simeq 3\rstar$ but there is evidence of higher velocities ($\gtrsim 9-10$~\kms) beyond this region.
In the vicinity of the star, the line widths are as high as $\simeq 10$~\kms, which implies a maximum turbulent velocity of 6~\kms{} or the existence of other physical mechanisms probably related to matter ejection that involve higher gas expansion velocities than expected.
HCN is rotationally and vibrationally out of LTE throughout the whole envelope.
It is surprising that a difference of about 1500~K in the rotational temperature at the photosphere is needed to explain the observations at 7.5 and $13-14~\mu$m.
Our analysis finds a total HCN column density that ranges from $\simeq 2.1\times 10^{18}$ to $3.5\times 10^{18}$~\cmm, an abundance with respect to H$_2$ of $3.5\times 10^{-5}$ to $1.3\times 10^{-4}$, and a $^{12}$C/$^{13}$C isotopic ratio of $\simeq 2.5$ throughout the whole envelope.
}

\maketitle

\section{Introduction}
\label{sec:introduction}

The structure, kinematics, and chemistry of circumstellar envelopes (CSEs) of the Asymptotic Giant Branch stars (AGBs) have been explored since the second half of the last century.
These variable stars return processed matter to the interstellar medium by a combination of the pulsation of their atmospheres, which creates the foundation of the CSE, and the formation of dust grains as a consequence of the gas cooling that subsequently occurs in this environment \citep*[e.g.,][]{goldreich_1976,bowen_1988,decin_2006}.
The development of these winds are tightly related to the chemical evolution of the expelled gas, which depends on the C/O abundance ratio \citep*[e.g.,][]{agundez_2020}.
Carbon rich stars (C-rich) are defined as those with CSEs where C/O~$>1$, the envelopes of oxygen rich stars (O-rich) show C/O~$<1$, and the S-type stars have CSEs where C/O~$\simeq 1$.
Most of what we know about these environments has been derived from the study of \irc{} \citep[the C-rich CSE around the star CW~Leo; e.g.,][]{keady_1988,keady_1993,cernicharo_2000,fonfria_2008,agundez_2012}, IK~Tau \citep[O-rich; e.g.,][]{velilla-prieto_2017,decin_2018a,decin_2018b}, or W~Aql \citep[S-type; e.g.,][]{danilovich_2014,ramstedt_2017,debeck_2020}, with mass-loss rates of $\simeq 2.7\times 10^{-5}$, $2.0\times 10^{-5}$, and $2.7\times 10^{-6}$~\mlr, respectively \citep{ramstedt_2014,guelin_2018}.

The envelopes of these high mass-loss rate stars are prime locations to search for new molecules, investigate the dust density distribution or look for gas structures \citep*[e.g.,][]{mauron_1999,cernicharo_2000,cernicharo_2015a,cernicharo_2015b,cernicharo_2019a,cernicharo_2019b,leao_2006,agundez_2008,debeck_2013,jeffers_2014,velilla-prieto_2017}.
Nevertheless, investigating the inner layers of the envelopes, where the gas is accelerated \citep[$r\simeq 1-30\rstar$; e.g.,][]{decin_2006,fonfria_2008}, remained hampered because of the increased density of gas and dust in this region and the demanding angular resolution requirements.
Recent observations of vibrationally excited lines in addition to the large capacities of ALMA and the VLTI are shedding light on this problem \citep*[e.g.,][]{fonfria_2008,fonfria_2019,cernicharo_2011,agundez_2012,vlemmings_2017,adam_2019,ohnaka_2019}.

Meanwhile, low mass-loss rate stars ($\sim 10^{-7}$~\mlr) are, typically, much brighter in the optical and the near-IR because of their thinner CSEs.
This allows us to directly observe the central stars and inner layers of their envelopes.
However, the complexity of the electronic molecular transitions has limited the community to use these observations to understand the kinematics and chemistry close to their central stars with few exceptions.
\citet{hinkle_1978} and \citet{hinkle_1979a,hinkle_1979b} analyzed the gas kinematics of the extended atmosphere of the O-rich star \rleo, with a mass-loss rate of $1.0\times 10^{-7}$~\mlr{} \citep{ramstedt_2014}.
However, the molecular emission and gas kinematics of the extended atmosphere, and even the first tens of stellar radii beyond, of no C-rich counterpart has been studied in detail as far as we know.
A step was taken in this direction with V~Oph, whose C$_2$H$_2$ absorption between 8 and 9~$\mu$m was studied with low spectral resolution interferometer observations \citep{ohnaka_2007} to roughly determine the abundance and excitation temperature of this molecule near the photosphere.

About $10-15\%$ of C-rich stars are J-type stars \citep{abia_2000,morgan_2003}.
These stars show a significant enhancement of the $^{13}$C abundance with respect to the main carbon isotopologue \citep[$^{12}$C/$^{13}$C~$\lesssim 15$;][]{abia_1997,ohnaka_1999}, compared to the typical ratio found for most C-rich stars \citep[R-, N-, and CH-type, $30\lesssim {}^{12}$C/$^{13}$C~$\lesssim 100$;][]{lambert_1986,keenan_1993,chen_2007}.
The Li abundance is usually higher than expected \citep[e.g.,][]{gordon_1971,abia_1997,hatzidimitriou_2003} and heavy elements associated with \textit{s}-processes are barely present in their atmospheres \citep[e.g.,][]{dominy_1985}.
Most of J-type stars are irregular or semi-regular \citep{abia_2000} and their mass-loss rates are usually low \citep{lorenz-martins_1996}.
The origins of J-type stars seem to be different to those of typical C-rich (N stars) and SC stars, i.e., those which are presumably evolving from S-type to C-rich \citep{abia_2020}.
Several hypotheses have been proposed to explain the peculiarities of these stars, such as that they are evolving independently to ordinary carbon stars \citep{lorenz-martins_1996} or they are the descendants of low-mass R stars \citep{abia_2003}.
Nevertheless, there is not a satisfactory explanation to their existence thus far.

\ycvn{} is a C-rich J-type semi-regular b (SRb) star \citep[e.g.,][]{abia_2000} that is located at $\simeq 310\pm 17$~pc from Earth \citep[Gaia Early Data Release 3 (EDR3);][]{gaia,gaia_edr3}.
Its mass-loss rate at this distance is $\simeq 1.4\times 10^{-7}$~\mlr{} based on the results of \citet{ramstedt_2014} ($\simeq 1.5\times 10^{-7}$~\mlr{} for a distance of 321~pc).
It was observed in the past many times at different spectral resolutions between the optical and millimeter ranges \citep*[e.g.,][]{lambert_1986,olofsson_1993a,olofsson_1993b,bachiller_1997,schilke_2003,yang_2004,neufeld_2011,schoier_2013,ramstedt_2014,massalkhi_2018b,massalkhi_2019}.
These observations revealed that the CSE of \ycvn{} displays molecular features from species such as CO, HCN, CN, CS, SiS, SiC$_2$, SiO, C$_2$, C$_3$, CH, and H$_2$O.
Low angular resolution observations ($\simeq 2\arcsec-3\arcsec$) of HCN($1-0$) and CN($1-0$) are available \citep{lindqvist_2000}.
The HCN emission is point-like due to the compact maser contributions displayed by the $J=1-0$ line \citep[e.g.,][]{dinh-v-trung_2000}.
In contrast, the CN brightness distribution, which is expected to encircle the HCN brightness distribution, is partially resolved, and shows a clumpy, asymmetric structure that peaks at a distance of $1\arcsec-2\arcsec$ from \ycvn{} and has a diameter of about $6\arcsec$.
However, its CSE has not been explored in detail and the gas kinematics and the physical and chemical conditions throughout the envelope are poorly known mostly due to the lack of sensitive observations of the gas acceleration region.

In this paper, we present the first high spectral resolution observations in the mid-IR of \ycvn{} carried out with the Echelon-cross-Echelle Spectrograph (EXES; \citealt{richter_2018}) mounted on the Stratospheric Observatory for Infrared Astronomy (SOFIA; \citealt{temi_2018}) and with the Texas Echellon-cross-Echelle Spectrograph (TEXES; \citealt{lacy_2002}) mounted on the Infrared Telescope Facility (IRTF), which is located on Maunakea (Hawaii).
The analysis of the ro-vibrational spectrum of HCN~\footnote{From now on, we will use HCN to refer to the main isotopologue of hydrogen cyanide (H$^{12}$CN). H$^{13}$CN will be explicitly included in the text when we refer to this particular isotopologue.} and H$^{13}$CN together with new data acquired in the millimeter range with the IRAM 30~m telescope, placed on Pico Veleta (Spain), has allowed us to describe the kinematic behavior of the gas, the physical conditions prevailing throughout the whole envelope, and the abundance distribution of these molecules.

\section{Observations}
\label{sec:observations}

\subsection{SOFIA/EXES}

\begin{figure*}
  \centering
  \includegraphics[width=\textwidth]{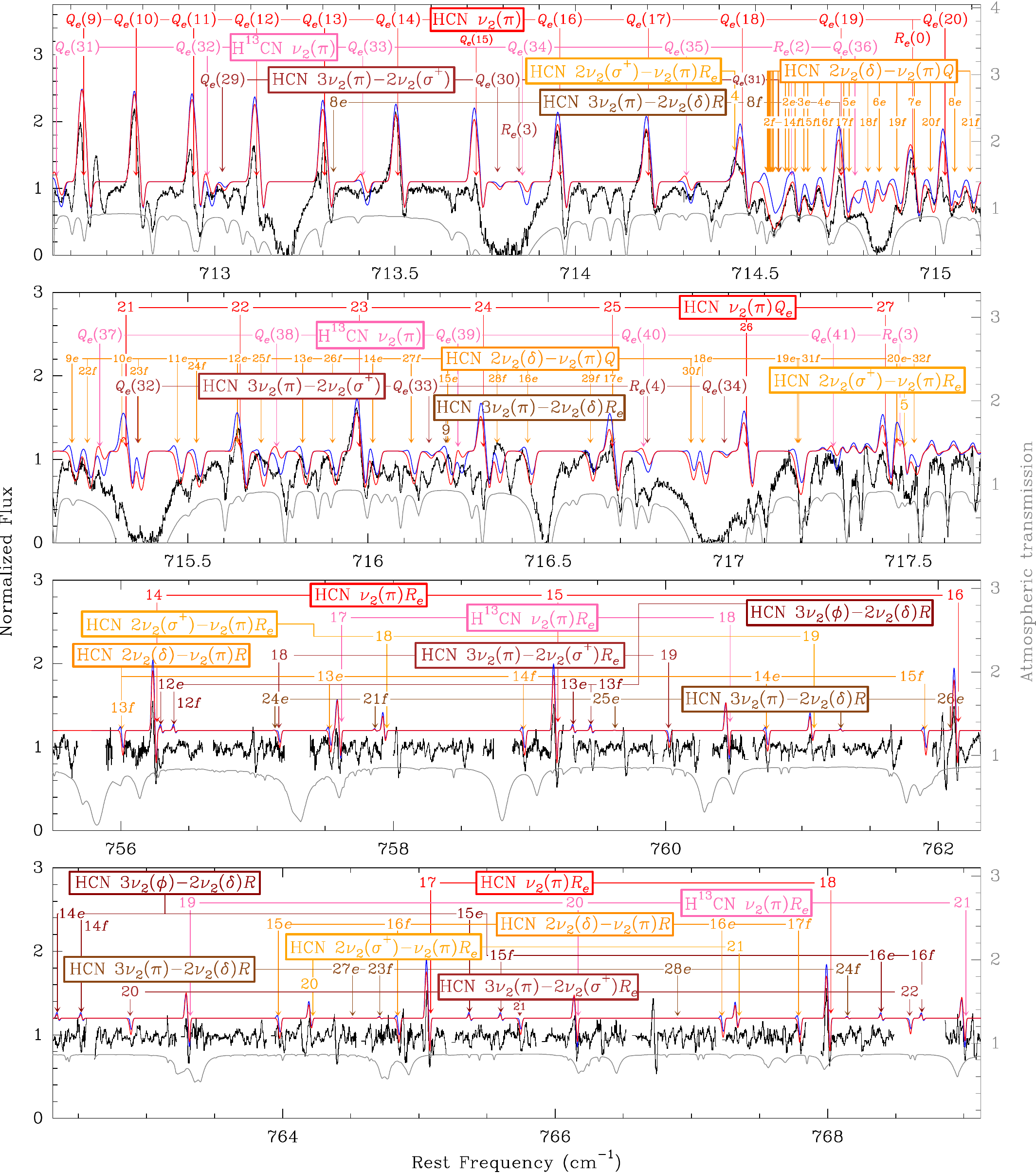}
  \caption{Observed spectrum of \ycvn{} in the spectral range $13-14$~$\mu$m (top panels: 14.0~$\mu$m, $712.55-717.70$~\cm, SOFIA/EXES; bottom panels: 13.1~$\mu$m, $755.50-769.12$~\cm, IRTF/TEXES).
    The spectra are corrected for the Doppler shift due to the relative movement between the star and the Earth at the time of the observations.
    The resolving power is $R\simeq 67,000$ at 14.0~$\mu$m and $\simeq 83,000$ at 13.1~$\mu$m, which provides us with a spectral resolution of $\simeq 4.4$ and 3.6~\kms, respectively.
    Different lines of several vibrational states of HCN (red, orange and brown) and H$^{13}$CN (pink) have been found in this spectrum.
    An ATRAN model of the atmospheric transmission is included in gray.
    The TEXES data have been cleaned from the atmospheric lines with a routine developed by J. H. Lacy.
    The red and blue curves are the synthetic spectra derived from our spherically symmetric and asymmetric models, respectively.
  }
  \label{fig:f1}
\end{figure*}

\begin{figure*}
  \centering
  \includegraphics[width=\textwidth]{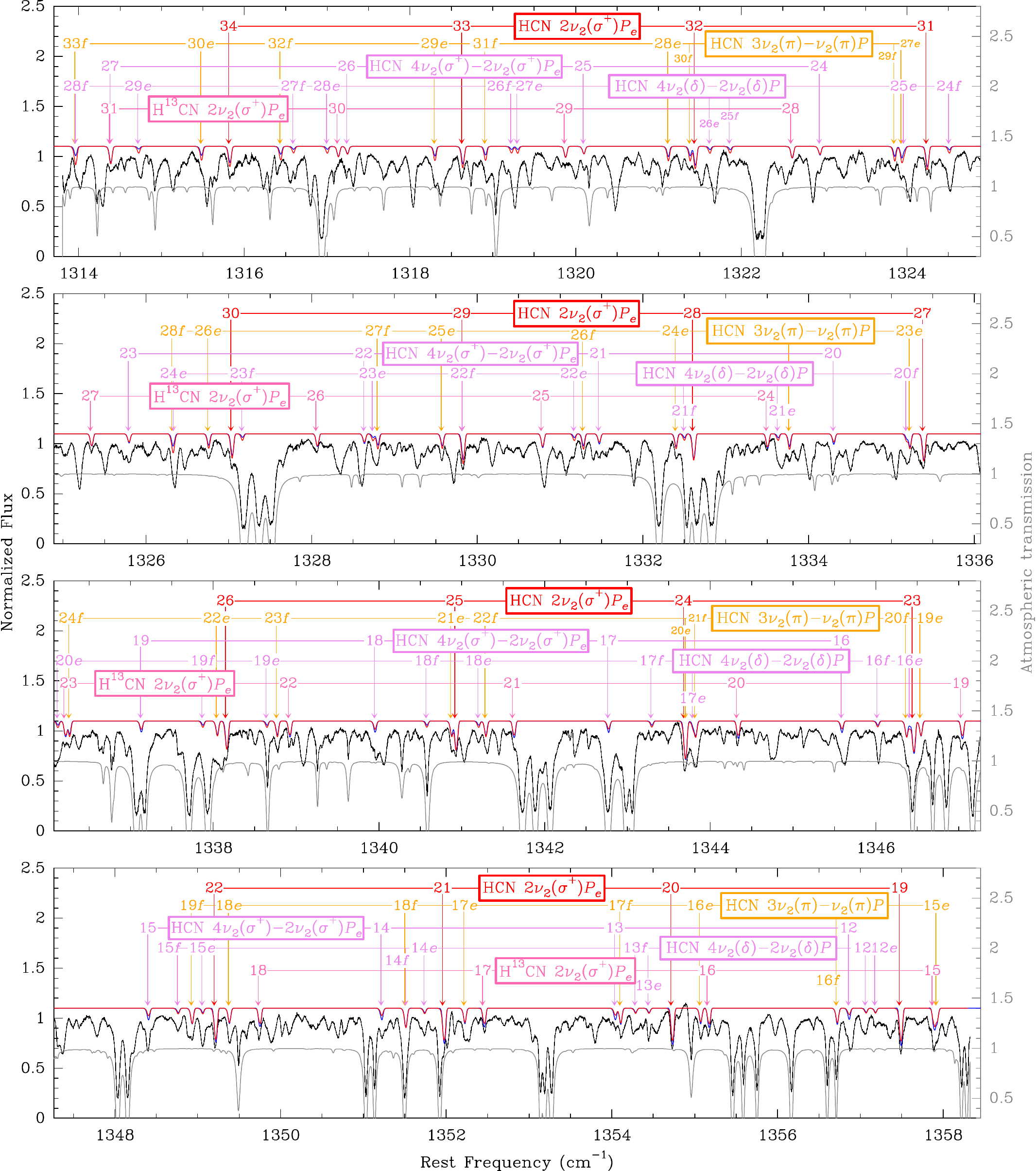}
  \caption{Observed spectrum of \ycvn{} at 7.5~$\mu$m ($1313.70-1358.45$~\cm) using SOFIA/EXES.
    The spectrum is corrected for the Doppler shift due to the star and Earth relative movement at the time of the observations.
    The resolving power is $R\simeq 60,000$ which provides us with a spectral resolution of $\simeq 5.0$~\kms.
    Different lines of several vibrational states of HCN (red, orange, violet) and H$^{13}$CN (pink) have been found in this spectrum.
    An ATRAN model of the atmospheric transmission is included in gray.
    The red and blue curves are the synthetic spectra derived from our spherically symmetric and asymmetric models, respectively.
  }
  \label{fig:f2}
\end{figure*}

\ycvn{} was observed with EXES on two successive SOFIA flights, which took place 17--18 Apr 2019 (UT).
The first run was conducted while SOFIA flew at an altitude of 12.8~km.
We focused on a central wavenumber of 1335 cm$^{-1}$ (7.5~$\mu$m) in the high-low instrument configuration.
We nodded \ycvn{} off of the slit to get to blank sky.
The slit was 2\farcs4 wide and 2\farcs0 long.
The full coverage of \ycvn's spectrum at this wavenumber and observing mode is 1313.7 to 1358.2  cm$^{-1}$ (7.362 to 7.612~$\mu$m).

The second observation of \ycvn{} was on the next flight at an altitude of 13.1~km in the high-medium mode. 
We used a 2\farcs4 wide slit with a length of 21\farcs0, so that nodding along the slit was possible.
We chose to center the observations around 714.70~cm$^{-1}$ (14.0~$\mu$m), which results in the covered spectral range $712.49-717.70$~cm$^{-1}$ ($13.93-14.04$~$\mu$m).

All EXES data were reduced using the Redux pipeline \citep{clarke_2015}.
The median spectral resolving power, $R=\lambda/\Delta\lambda$, for the high-low configuration at 7.5~$\mu$m was empirically determined to be about 60,000 by fitting several narrow, optically thin telluric features with a Gaussian.
This resolving power means a velocity resolution of 5.0~\kms.
An $R$ of 67,000 (4.4~\kms) was found for the high--medium setting at 14.0~$\mu$m.
Customized procedures were written for the 14.0~$\mu$m observations to extract an extra order at the high frequency edge of the covered range, which only contained our negative beam.

The observations show a forest of molecular lines in the covered spectral range dense enough to make difficult finding a clean baseline.
A careful inspection of the spectra suggests that it is affected by a low amplitude, long wavelength ripple that has been removed by dividing the data with a cubic spline fit.
We have not found periodic features or artifacts affecting the line profiles.
The resulting normalized spectra together with the identified HCN and H$^{13}$CN lines are shown in Figures~\ref{fig:f1} and \ref{fig:f2} (see a brief description of HCN in appendix~\ref{sec:intro.HCN} and a vibrational energy diagram showing the vibrational transitions observed in this work in Figure~\ref{fig:fa1}).
The lines of other molecules are not identified in Figures~\ref{fig:f1} and \ref{fig:f2} for the sake of clarity.
We estimate a noise RMS of $\simeq 0.5$\% of the continuum at 7.5~$\mu$m and of $\simeq 2.0$\% at 14.0~$\mu$m.
We have adopted a systemic velocity in the LSR system of +22.0~\kms{} \citep{massalkhi_2018b}, which implies radial velocities of \ycvn{} with respect to us at the time of the observations of +24.8 and +25.1~\kms{} at 7.5 and 14.0~$\mu$m, respectively.

\clearpage

\subsection{IRTF/TEXES}

Y~CVn was also observed with the Texas Echelon--Cross--Echelle Spectrograph (TEXES; \citealp{lacy_2002}) at NASA's IRTF on Maunakea.
These observations took place on 14 Feb 2019 (UT) and were carried out in TEXES' high-low mode.
They have a spectral coverage of 754.66 to 769.18 cm$^{-1}$ (13.00 to 13.25~$\mu$m).
The data were reduced following the methods described in \citet{lacy_2002}.
The value of $R$ was observationally determined from the data to be 83,000 (or a spectral resolution of 3.6~\kms).
We estimate a noise RMS of $\simeq 2$\% of the continuum.
The relative radial velocity of \ycvn{} during the observation is estimated to be +3.6~\kms.

After doing the standard data reduction, we normalized the spectrum and removed the weak and medium strength telluric features by dividing the observed data by an atmospheric transmission model derived from a routine written for TEXES by J.~H.~Lacy.
It relies on the HIgh-resolution TRANsmission molecular absorption database (HITRAN; \citealt{hitran}) and balloon data to account for temperature and humidity against altitude from the Integrated Global Radiosonde Archive  project \citep{durre_2008}\footnote{This project is run by the National Centers for Environmental Information which is a part of the National Oceanic and Atmospheric Administration.}.
This code is still under development and it cannot be applied to the observations acquired with SOFIA/EXES.
Nevertheless, the results of using Lacy's model are better than what we got by dividing the observations by a telluric calibrator.
The spectral ranges blocked by strong telluric lines could not be recovered after the spectra division and they appear as gaps in Figure~\ref{fig:f1}.
The spectral ranges affected by weaker telluric features that were successfully removed show a higher noise but allow us to use molecular lines seriously contaminated otherwise.

\subsection{IRAM 30~m telescope}

\begin{figure}
  \centering
  \includegraphics[width=0.475\textwidth]{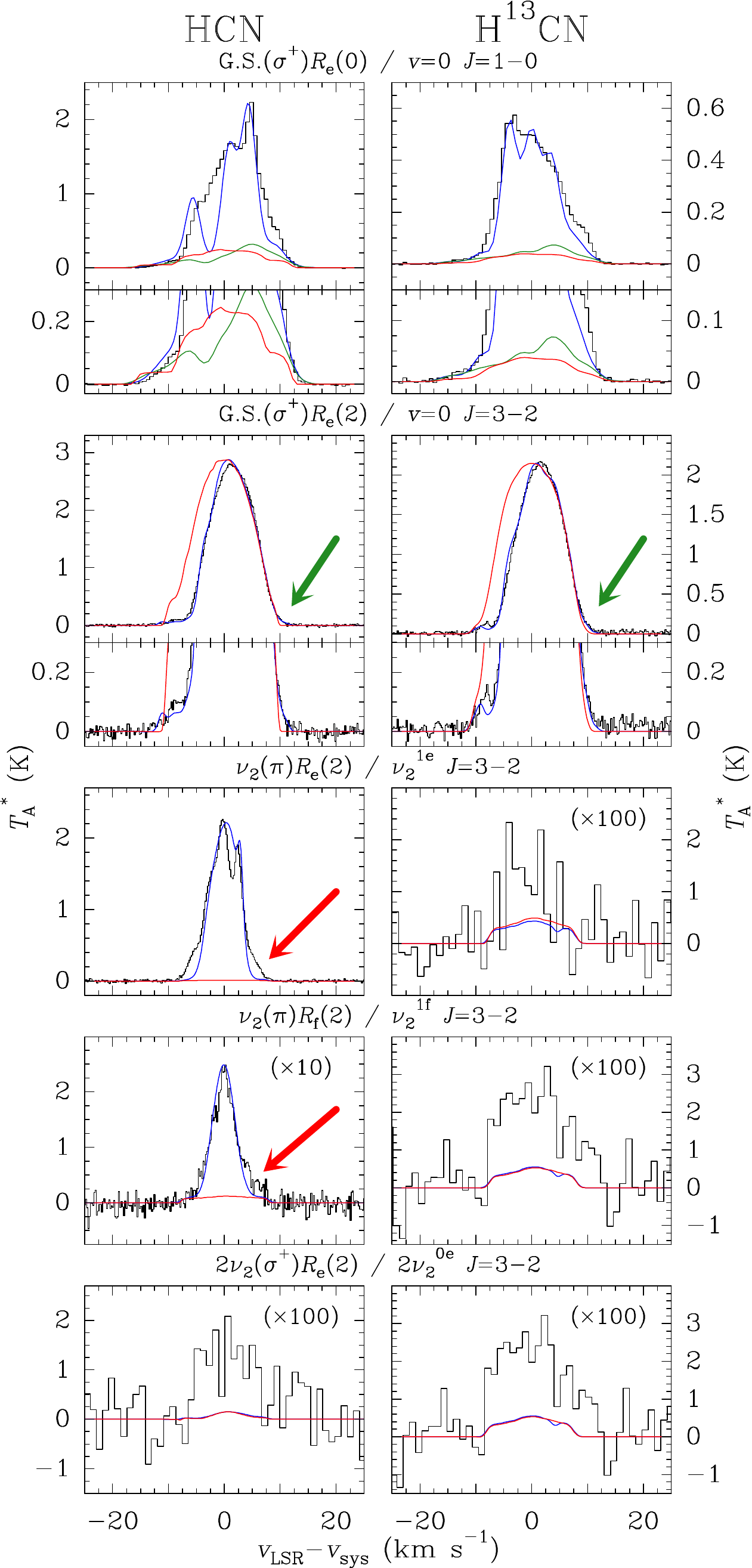}
  \caption{Observed HCN and H$^{13}$CN lines of \ycvn{} with the IRAM 30~m telescope (black histograms).
    The red and blue curves are the synthetic lines derived from the spherically symmetric (red) and asymmetric (including maser emission; blue) envelope models.
    The red and green arrows point to red-shifted wings that cannot be reproduced with our spherically symmetric envelope model.
    The small insets in the boxes with the $v=0$ $J=1-0$ and $3-2$ lines zoom in to show the line profiles close to the baseline (Section~\ref{sec:molecular.emission}).
    The green profiles for these lines have been calculated with the asymmetric envelope model without considering maser emission (Section~\ref{sec:discussion.velocity}).
    The H$^{13}$CN lines $\nu_2^{1f}$ and $2\nu_2^{0e}$ are strongly overlapped and the blended line has been included twice (lower right panels).
    The lines of the vibrational states $2\nu_2^2$ and other with higher energies are below the detection limit.
    In this Figure we have used two notations to name the pure rotational lines (see Appendix~\ref{sec:spectroscopic.data}).
}
  \label{fig:f3}
\end{figure}

The \ycvn{} observations were taken in 2017 and the first half of 2019 as part of projects 164-15, 155-16, 046-17, and 048-17 (Figure~\ref{fig:f3}).
We used the EMIR receivers E090 and E230 \citep{carter_2012} with the FTS200 backends, which provided us with a total bandwidth of 15.7~GHz and a channel width of 0.2~MHz.
The weather was usually good during the observing campaigns with atmospheric opacities below 0.25 at 1~mm and up to 0.05 at 3~mm.
The system temperatures were usually below 290 and 75~K, respectively.
The noise RMS levels are typically 12 and 7~mK ($T_A^*$; 100 and 55~mJy) for the lines of HCN and H$^{13}$CN at 1~mm, respectively, and 2.3~mK ($T_A^*$; 14~mJy) for the lines at 3~mm.
We used the wobbler switching technique wobbling the secondary mirror at a frequency of 0.5~Hz and with a throw of 90\arcsec.
This technique produces very flat baselines that favors the detection of weak lines.
We estimate the calibration error to be $\simeq 20\%$.
The $J=3-2$ lines of HCN and H$^{13}$CN, observed in different setups, have been made compatible by comparing the SiC$_2$ lines $12_{0,12}-11_{0,11}$ and $11_{2,10}-10_{2,9}$, also observed in different setups, whose intensities are expected to be nearly the same \citep{cernicharo_2012}.

\subsection{\textit{ISO}/SWS and LWS}

These data sets have been complemented with previous low spectral resolution observations acquired with the Infrared Space Observatory (\textit{ISO}) \citep{hony_2002,yang_2004}, which operated from 1995 to 1998, to derive the overall properties of the infrared continuum source of \ycvn.
This facility was able to take infrared spectra with the Short-Wavelength Spectrometer (SWS; $2.5-45~\mu$m) and the Long-Wavelength Spectrometer (LWS; $45-200~\mu$m).
\ycvn{} was observed with both spectrographs (TDT numbers: 16000926 and 19500730) and the grating modes S01 and L01, which gave effective resolving powers of 500 at $35~\mu$m and $200-300$ above $45~\mu$m, respectively \citep{degraauw_1996,yang_2004}.
The flux calibration of this spectrum has been checked against observations of different catalogs \citep[Two Micron All Sky Survey (2MASS), AKARI/IRC All-Sky Survey (AKARI), Cosmic Background Explorer Diffuse Infrared Background Experiment (DIRBE), Infrared Astronomical Satellite (IRAS), Widefield Infrared Survey Explorer (WISE), and the Unblurred coadds of the WISE imaging (unWISE) Point Source Catalogs;][]{neugebauer_1984,smith_2004,skrutskie_2006,ishihara_2010,wright_2010,lang_2014}.
The deviation is within the estimated uncertainties of 20\% for the \textit{ISO} data but we have scaled this spectrum to get the best agreement.

\section{Modeling}
\label{sec:modeling}

\subsection{Envelope model}
\label{sec:envelope.model}

The envelope has been divided into three different regions (I, II, and III) with inner radii $R_\subscript{\textsc{I}}=1\rstar$, \rdin, and \rdout{} (see Table~\ref{tab:table1}), where changes in important quantities such as the gas expansion velocity field and the formation of dust grains may occur.
The abundances of HCN and H$^{13}$CN have been considered as zero beyond the photodissociation radius, \rph, which is a free parameter (Table~\ref{tab:table2}).
The properties of the \ycvn{} CSE are poorly known due to the limited work that has focused on this low mass-loss rate star.
Nevertheless, some parameters have been previously well determined and they have been adopted as fixed in this work (Table~\ref{tab:table1}; non-free parameters).
Others have been taken from the literature as initial guesses for our model.
Necessary quantities absent in the literature have been adopted to be equal to those derived from other sources.

\begin{table*}
\caption{Non-free parameters involved in the fits\label{tab:table1}}
\centering
\begin{tabular}{cccc}
\hline\hline
Parameter & Units & Value & Refs./Comments\\
          &       &       & \\
\hline
$D$                                              & pc                         & $310\pm 17$                  & 1 \\
$\alpha_\star$                                    & mas                       & $7.0\pm 0.4$                  & 2 \\
\rstar                                           & cm                         & $(3.3\pm 0.4)\times 10^{13}$  & \\
$\dot M$                     & M$_\odot$~yr$^{-1}$         & $(1.4\pm 0.4)\times 10^{-7}$  & 3,S \\
\rdin                                            & \rstar                     & $3.2$                         &  A \\
\rdout                                           & \rstar                     & $200$                         &  A\\
$\tk(r)$                                         & K                          & $T_\star(\rstar/r)^{0.675}$    &  4   \\
$T_\subscript{dust}(r)$                           & K                          & $T_{d0}(\rstar/r)^{0.4}$       & 5  \\
$n_\subscript{dust}(\rstar\le r<\rdin)$           & cm$^{-3}$                   & 0.0                          &  \\
$f_\subscript{AC}(\rdin\le r<\rdout)$             & \%                         & $0$                           & A\\
$f_\subscript{AC}(r\ge\rdout)$                    & \%                         & $100$                         & A\\
$f_\subscript{SiC}(\rdin\le r<\rdout)$            & \%                         & $100$                         & A\\
$f_\subscript{SiC}(r\ge\rdout)$                   & \%                         & $0$                           & A\\
$\Delta v(r\ge\rdin)$                            & \kms                       & $1.0\pm 0.2$                  & 6 \\
$x_\subscript{He}$                                &                            & $0.17$                        & 7\\
\hline                                           
\end{tabular}
\tablefoot{
The non-free parameters were fixed throughout the modeling process.
$D$: distance to the star; 
$\alpha_\star$: angular stellar radius;
$R_\star$: linear stellar radius (inner radius of Region I);
$\dot M$: mass-loss rate; 
\rdin{} and \rdout: position of the inner boundaries of Regions II and III ($R_\subscript{\textsc{I}}=1\rstar$);
$\tk(r)$: gas kinetic temperature ($T_\star$ is a free parameter; see Table~\ref{tab:table2});
$T_\subscript{dust}$: temperature of dust grains ($T_{d0}$ is a free parameter; see Table~\ref{tab:table2});
$n_\subscript{dust}(r)$: density of dust grains ($\propto r^{-2}$);
$f_\subscript{X}$: fraction of dust grains composed of material X (AC: amorphous carbon; SiC: silicon carbide);
$\Delta v$: line width;
$x_\subscript{He}$: He abundance with respect to H$_2$.
References: 
(1) Gaia EDR3 \citep{gaia_edr3}
(2) \citet{ragland_2006}
(3) \citet{ramstedt_2014}
(4) \citet{guelin_2018}
(5) \citet{sopka_1985}, $p=1$
(6) \citet{keady_1988},
(7) \citet{asplund_2009}.
Comments:
A: parameter assumed after many tries,
S: scaled to account for the new distance.
}
\end{table*}

\begin{table*}
\caption{Free parameters involved in the fits\label{tab:table2}}
\centering
\small
\begin{tabular}{ccccccc}
\hline\hline
Parameter & Units & Continuum & \multicolumn{3}{c}{High Resolution Observations} & Comments\\
          &       &           & $7.5~\mu$m  & $13-14~\mu$m & mm & \\
\hline                                           
\rph(HCN)                                       & \rstar                     & ---            & ---          &  ---           &  $168\pm 22$ &   \\ 
\rph(H$^{13}$CN)                    & \rstar                     & ---            & ---          &  ---           &  $155\pm 20$ &   \\ 
$T_\star$                                         & K                          & $2910\pm 70$   & ---          & ---           & ---        &  \\
$T_{d0}$                                         & K                           & $1280\pm{}^{90}_{100}$  & ---          & ---           & ---        &  \\
$n_\subscript{dust}(\rdin)$                       & $\times 10^{-5}$~cm$^{-3}$  & $1.30\pm{}^{0.22}_{0.30}$   & ---          & ---           & ---        &  \\
$n_\subscript{dust}(\rdout)$                      & $\times 10^{-8}$~cm$^{-3}$  & $6.2\pm 0.9$   & ---          & ---           & ---        &  \\
$v_\subscript{exp}(\rstar)$                       & \kms                       & ---            & \multicolumn{2}{c}{$0.50\pm{}^{0.05}_{0.03}$} & ---        &  \\
$v_\subscript{exp}(r\ge\rdin)$                      & \kms                       & ---            & \multicolumn{2}{c}{$8.00\pm 0.11$}   & ---        &  \\
$\Delta v(\rstar)$                               & \kms                       & ---            & \multicolumn{2}{c}{$10.0\pm{}^{0.8}_{1.6}$}  & ---        &  \\
\trot(\rstar)                                    & K                          & ---            & $2910\pm{}^{280}_{230}$ & $1200\pm{}^{230}_{140}$    & ---        &  \\ 
\trot(\rdin)                                     & K                          & ---            & $900\pm 60$      & $700\pm 50$     & ---        &  \\ 
\trot(\rdout)                                    & K                          & ---            & ---           & ---          &  $79\pm 11$   &  \\ 
$T_\subscript{vib}[\nu_2(\pi)-\textnormal{GS}(\sigma^+)](\rstar)$                       & K                          & ---            & \multicolumn{2}{c}{$1200\pm{}^{50}_{40}$} & ---        &  \\ 
$T_\subscript{vib}[\nu_2(\pi)-\textnormal{GS}(\sigma^+)](\rdin)$                        & K                          & ---            & $330\pm{}^{13}_{19}$      & $360.0\pm 1.4$     & ---        &   \\ 
$T_\subscript{vib}[\nu_2(\pi)-\textnormal{GS}(\sigma^+)](\rdout)$                       & K                          & ---            & $30\pm{}^{170}_{30}$       & $70.0\pm 2.1$      & ---        & I \\ 
$T_\subscript{vib}[2\nu_2(\sigma^+)-\nu_2(\pi)](\rstar)$                       & K                          & ---            & $1200\pm{}^{100}_{70}$     &   $1500\pm{}^{120}_{90}$         & ---        &  \\ 
$T_\subscript{vib}[2\nu_2(\sigma^+)-\nu_2(\pi)](\rdin)$                        & K                          & ---            & $330\pm 30$      &   $400\pm 40$          & ---        &   \\ 
$T_\subscript{vib}[2\nu_2(\sigma^+)-\nu_2(\pi)](\rdout)$                       & K                          & ---            & $30$       &   $70$           & ---        & I,F \\ 
$T_\subscript{vib}[2\nu_2(\delta)-2\nu_2(\sigma^+)](\rstar)$                       & K                          & ---            &        \tk(\rstar)     &    $50\pm{}^{12}_{8}$        & ---        & F,S \\ 
$T_\subscript{vib}[2\nu_2(\delta)-2\nu_2(\sigma^+)](\rdin)$                        & K                          & ---            &        \tk(\rdin)      &    $20\pm{}^{10}_{4}$       & ---        & F,S  \\ 
$T_\subscript{vib}[2\nu_2(\delta)-2\nu_2(\sigma^+)](\rdout)$                       & K                          & ---            &        \tk(\rdout)      &   $10$       & ---        & I,F,S \\ 
$T_\subscript{vib}[\nu_3(\sigma^+)-2\nu_2(\delta)](\rstar)$                       & K                          & ---            & $1200\pm{}^{50}_{40}$     &   $1500\pm{}^{420}_{230}$         & ---        &  \\ 
$T_\subscript{vib}[\nu_3(\sigma^+)-2\nu_2(\delta)](\rdin)$                        & K                          & ---            & $330\pm{}^{13}_{19}$      &   $450\pm{}^{400}_{210} $          & ---        &   \\ 
$T_\subscript{vib}[\nu_3(\sigma^+)-2\nu_2(\delta)](\rdout)$                       & K                          & ---            & $30$       &      $70$            & ---        & I,F \\ 
$x(\rstar)$                                     & $\times 10^{-4}$           & ---            &  $1.30\pm{}^{0.05}_{0.08}$    &  $0.35\pm 0.05$   & ---         &  \\
$x(\rdin)$                                      & $\times 10^{-4}$           & ---            &  \multicolumn{2}{c}{$1.30\pm{}^{0.05}_{0.08}$}& ---         &   \\ 
$x(r\ge\rdout)$                                     & $\times 10^{-4}$            & ---            &    ---       &  ---         & $1.3\pm 0.7$   &  \\ 
$^{12}$C/$^{13}$C           &                            & ---            & \multicolumn{2}{c}{$2.50\pm 0.15$}    & $2.5\pm 1.0$         &  \\
\hline
\multicolumn{7}{c}{Other useful derived parameters}\\
\hline
$\tau_\subscript{dust}(11.3~\mu\textnormal{m})$  & $\times 10^{-3}$             & $3.3\pm{}^{0.6}_{0.7}$  & ---          &  ---         & ---                  &  \\
$N_\subscript{col}(\textnormal{HCN})$            & $\times 10^{18}$~cm$^{-2}$   &  ---            & \multicolumn{3}{c}{$2.1\pm 1.1$}           &  V\\
$N_\subscript{col}(\textnormal{H}^{13}\textnormal{CN})$    & $\times 10^{17}$~cm$^{-2}$   & ---   &  \multicolumn{3}{c}{$8\pm 4$}          &  V \\
\hline
\end{tabular}
\tablefoot{
A few initially free parameters were fixed at some point during the line modeling due to the model insensitivity to them (see below).
These parameters do not show uncertainties.
The uncertainties are $1\sigma$ errors.
All the molecular parameters are for HCN and are valid for H$^{13}$CN unless it is indicated.
\rph: photodissociation radius;
$T_\star$: stellar effective temperature;
$T_{d0}$: temperature of dust grains at \rdin;
$n_\subscript{dust}(r)$: density of dust grains ($\propto r^{-2}$);
$v_\subscript{exp}$: gas expansion velocity (continuous function with a linear dependence on $r$ in each Region);
$\Delta v$: line width;
\trot: rotational temperature;
$T_\subscript{vib}$: vibrational temperature for consecutive bands and energy differences higher than 100~K (for lower differences $T_\subscript{vib}=\tk$);
$x$: HCN abundance with respect to H$_2$ (continuous function with a linear dependence on $r$ in each Region);
$\tau_\subscript{dust}$: dust optical depth along the line-of-sight;
$N_\subscript{col}(\textnormal{HCN})$ and $N_\subscript{col}(\textnormal{H}^{13}\textnormal{CN})$: column densities of HCN and H$^{13}$CN;
$^{12}$C/$^{13}$C: isotopic ratio.
Comments:
  I: model insensitive to at least one of these parameters;
  F: at least one of the parameters fixed during the calculations;
  S: these parameters might be affected by spectroscopic inaccuracies or an incomplete envelope model (see Section~\ref{sec:discussion.temperatures});
  V: average of the results derived from the fits to all the molecular lines.
}
\end{table*}

A terminal gas expansion velocity of $8-9$~\kms{} \citep{izumiura_1995,ramstedt_2014} along with a turbulent line width of 1~\kms{} as in the C-rich AGB star \irc{} \citep{fonfria_2008,agundez_2012} throughout the whole envelope, and a stellar effective temperature of 2760~K \citep{bergeat_2001} have been initially considered.
We have adopted the gas kinetic temperature dependence in the CSE of \ycvn{} as a function of the distance from the central star, which to first-order approximation (the temperature at the photosphere of \ycvn{} is a free parameter) is derived from the outer envelope of \irc{} as it is reasonably well known \citep{guelin_2018}.
This region of the CSE of \irc{} is as rarefied as the envelope of \ycvn, and both envelopes are chemically similar (see Section~\ref{sec:discussion.temperatures} for a discussion about this assumption).
We have considered vibrational temperatures between vibrational states affected by $\ell$-doubling (e.g., $2\nu_2(\delta)-2\nu_2(\sigma^+)$, $\Delta E_\subscript{vib}\simeq 30$~K) to be equal to the kinetic temperature unless different values are necessary.
The gas density has been derived from the mass conservation law throughout the whole CSE, i.e., $n_\subscript{gas}\propto r^{-2}v_\subscript{exp}^{-1}$, where $v_\subscript{exp}$ is the gas expansion velocity and it can depend on the distance from the star.

The dust grains have been assumed to be composed of amorphous carbon (AC) and silicon carbide (SiC), as these materials explain most of the continuum emission at wavelengths below $\simeq 30~\mu$m in C-rich AGB stars \citep*[e.g.,][]{keady_1988,hony_2002,swamy_2005}.
We have considered that carbon-bearing molecules (probably C$_2$ and C$_3$) are able to condense into the carbonaceous material that compose the dust grains \citep*[see Figure~5 of][]{agundez_2020} at the starting point of $\simeq 2\rstar$ from the center of \ycvn.
Solid state SiC is expected to form due to the condensation of Si$_n$C$_m$ molecules on the carbonaceous seeds at the position of the envelope where the kinetic temperature is $\simeq 1300$~K, i.e., at $\simeq 3\rstar$ from the center of the star \citep{gobrecht_2017,agundez_2020}.
The density and temperature of dust grains are assumed to follow the power laws $\propto r^{-2}$, typical of isotropic expansions, and $\propto r^{-0.4}$ \citep{sopka_1985}, respectively.

Dust grains are expected to be small in expanding, rarefied winds where the low gas density hampers the dust coagulation  \citep*[e.g.,][]{hirashita_2000}.
We have thus assumed small, spherical dust grains with a radius of $0.05~\mu$m and a density of 2.5~g~cm$^{-3}$.
We have not included scattering because it can be considered to be negligible at wavelengths longer than 5~$\mu$m for dust grains with a diameter $\lesssim 1~\mu$m.

Spherical symmetry has been initially assumed for all the physical and chemical quantities.

\subsection{The ro-vibrational diagram}
\label{sec:boltzmann}

A molecular ro-vibrational line produced by an isotropically expanding envelope usually shows a P-Cygni profile, which comprises an absorption component and an emission one typically overlapped.
The optical depth of the absorption component of an observed line, which is produced by the molecules in front of the continuum source (the star and the surrounding dust), can be described with this formula (see Appendix~\ref{sec:rovib.diagram}):
\begin{equation}
  \label{eq:rovib.diagram}
  \ln\left[\frac{W\nu^28\pi c}{A_{ul}g_u N_\subscript{col,0}}\right]\simeq
  y_0-\frac{hcE_\subscript{rot,low}}{k_\subscript{B}T_\subscript{rot}},
\end{equation}
where $W$ is the optical depth associated to the absorption component integrated over the frequency (\cm), $\nu$ is the rest frequency (\cm), $E_\subscript{rot,low}$ the rotational energy of the lower level involved in the transition, and $T_\subscript{rot}$ the rotational temperature of the lower vibrational state.
$N_{\subscript{col},0}$ is an arbitrary scaling factor to get a dimensionless argument for the logarithm.
Eq.~\ref{eq:rovib.diagram} holds for all the lines of a given ro-vibrational band.
We define the quantity $y_0$ as:
\begin{equation}
  \label{eq:y-intercept}
y_0=\ln\left[\frac{N_\subscript{col}}{N_\subscript{col,0}Z}\right]-\ln\left[\frac{e^{hcE_\subscript{vib,low}/k_\subscript{B}T_{\subscript{vib},l}}}{1-e^{-hc\bar\nu/k_\subscript{B}T_{\subscript{vib},ul}}}\right],
\end{equation}
where $N_\subscript{col}$ is the column density, $Z$ the total partition function ($Z\simeq Z_\subscript{rot}Z_\subscript{vib}$), $\bar\nu$ is the mean frequency of the considered lines, $E_\subscript{vib,low}$ the vibrational energy of the lower vibrational state, and $T_{\subscript{vib},l}$ and $T_{\subscript{vib},ul}$ describe the populations of the lower vibrational state and the upper one with respect to the lower via Boltzmann factors, respectively.

Fitting the data derived from all the lines in a band with Eq.~\ref{eq:rovib.diagram} we can estimate the rotational temperature of this band from the slope and the column density from the $y$-intercept, $y_0$ (Eq.~\ref{eq:y-intercept}).
Note that the data sets related to different hot bands or overtones display $y$-intercepts offset with respect to that of the fundamental bands due to the effect of the vibrational temperatures.

\subsection{Radiative transfer code}
\label{sec:code}

The ro-vibrational diagram gives rough estimates of the molecular excitation and the column density of a given molecule but a more detailed description of the envelope requires the use of more sophisticated methods.
We have analyzed the continuum and the molecular spectra with the aid of the radiation transfer code developed by \citet{fonfria_2008} to model spherically symmetric CSEs (1D) and improved by \citet{fonfria_2014} to deal with asymmetric envelopes (3D).
This code solves the radiation transfer problem in a circumstellar envelope composed of molecular gas, dust grains and a central continuum source.
More information about it, the adopted spectroscopic data and optical properties of dust grains, as well as of the uncertainties of the parameters can be found in the Appendix~\ref{sec:appendix.code} and in the papers previously cited.

All the lines of each molecule have been fitted simultaneously.
The success of the fit has been checked by eye.
The high line density in the spectra (Figures~\ref{fig:f1} and \ref{fig:f2}), which is frequently seen as a partial contamination of the line to be fitted by other features, together with the high variety of intensities of the observed HCN lines prevented us to use an automatic process based on the minimization of the $\chi^2$ function.
The general agreement between the automatically calculated best fit and the observations was always poorer.
The number of free parameters involved in the fitting process is usually high (see Table~\ref{tab:table2}) but many lines are fitted at the same time.
Moreover, the observed lines can be gathered in different groups with particular properties (vibrational states involved or formation region, for instance) and usually only a few parameters directly affect each group of lines while the rest of them have a limited impact.

\section{Results}
\label{sec:results}

\subsection{The continuum emission}
\label{sec:continuum}

\begin{figure}
  \centering
  \includegraphics[width=0.475\textwidth]{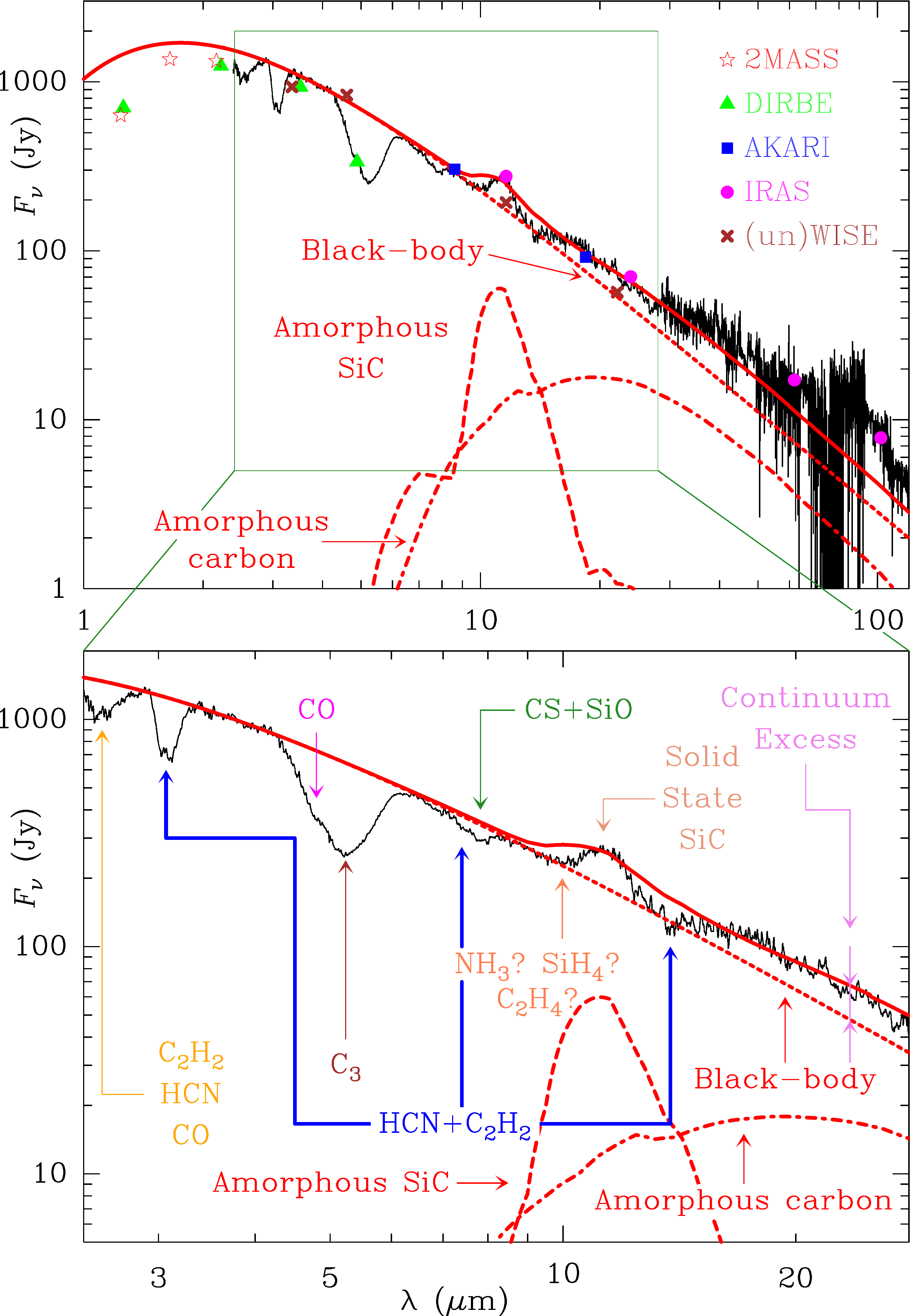}
  \caption{Continuum of \ycvn{} as observed with \textit{ISO} (black histogram).
    The red solid curve is the fit to the observations after the removal of the molecular bands (Table~\ref{tab:table2}).
    The red dotted curve is the stellar contribution, the red dashed curve the contribution of SiC, and the dotted-dashed one the contribution of amorphous carbon.
    The red stars, green triangles, blue squares, and magenta circles in the upper panel are photometric measures of the continuum available in different catalogs (2MASS, DIRBE, AKARI, IRAS, and WISE --- the WISE fluxes at 3.35 and 4.60~$\mu$m are clear outliers and they have been discarded to use the unWISE values instead).
    The band identifications in the lower panel are in agreement with previous observations of \ycvn{} and \irc{} \citep{cernicharo_1999,yang_2004}.}
  \label{fig:f4}
\end{figure}

The spectral energy distribution (SED) of \ycvn{} is single-peaked and the maximum emission is reached at $\simeq 1.9~\mu$m (Figure~\ref{fig:f4}).
It decreases monotonically with increasing wavelength except for a few molecular bands.
Considering that the mass-loss rate of this star is $\sim 10^{-7}$~M$_\odot$~yr$^{-1}$ and that the gas-to-dust ratio is $\simeq 500$ \citep{schoier_2001,massalkhi_2018b,massalkhi_2019}, the SED described by the \textit{ISO} observations is very probably dominated by the stellar continuum emission at short wavelengths.

The continuum has been fitted with our code in the wavelength range $\simeq 2.5-26~\mu$m.
We have taken parameters derived from \irc{} \citep*[e.g.,][]{fonfria_2008,fonfria_2015} as initial guesses.
No gas-phase molecular species has been considered in this fit and the molecular bands have not been reproduced.
The results can be found in Figure~\ref{fig:f4} and the derived parameters in Table~\ref{tab:table2}.
The continuum emission from 2.5 to $9~\mu$m is compatible with the emission of a black-body with a temperature of about 2900~K.
The clear SiC solid state band at 11.3~$\mu$m reveals the presence of dust grains composed of this material.
Below 2.5~$\mu$m, the synthetic continuum departs from the observations.
This disagreement is likely a direct consequence of the strong photospheric C$_2$ and CN absorption bands present below $2~\mu$m in the near-IR and the optical ranges typical of J-type stars \citep[e.g.,][]{barnbaum_1994,ohnaka_1999,massalkhi_2018a} that have not been included in our model.

The SiC solid state band is only slightly sensitive to temperature changes and the SiC absorption is comparatively negligible at any other wavelength.
However, the observed and synthetic SEDs are in better agreement when the condensation temperature is around 1000~K than when it is of a few hundreds of K.
Fixing the SiC condensation radius to a physically acceptable value is therefore necessary to propose an envelope model.
The addition of AC to the model with a dust temperature of the order of 1000~K produces a bump at $8-10~\mu$m that makes the model deviate significantly from the \textit{ISO} observations.
The only scenario that we have found to satisfactorily explain the overall emission beyond $2.5~\mu$m considers dust grains made of amorphous SiC with a contribution of AC at a temperature of a few hundred K.

Our results suggest that it is not possible to find a model that accounts for the emission excess beyond 15~$\mu$m without considering an opacity increase directly related to AC from $\simeq 100\rstar$ to $500\rstar$.
The best fit is achieved with such an increase located at $\simeq 200\rstar$, where the density of dust grains is enhanced by a factor of 15 with respect to the density resulting from a constant dust mass-loss rate.
Further opacity increases related to colder dust may be necessary to explain the continuum around 100~$\mu$m, which could be related to the extended dusty shell with an inner radius of $\simeq 3^\prime$ discovered long ago with \textit{IRAS} \citep{young_1993,izumiura_1996,libert_2007,cox_2012a,cox_2012b}.

\subsection{The molecular emission}
\label{sec:molecular.emission}  

The observed infrared spectra contain a forest of lines of different molecules.
In the $13-14$~$\mu$m range, the spectrum is dominated by broad features with strong emission components and weaker absorptions, typical of P-Cygni profiles coming from molecules with a high vibrational temperature.
There are also weaker lines showing P-Cygni profiles and other lines only in absorption.
Around 7.5~$\mu$m the spectrum is crowded with absorption lines.

Most of the lines in both spectral ranges come from HCN, \acet, and CS, in addition to their $^{13}$C-substituted isotopologues, as expected to happen in a J-type star such as \ycvn{} \citep*[$^{12}$C/$^{13}$C~$\simeq 3$; e.g.,][]{abia_2017}.
The observations contain high-$J$ CS lines of the $v=2-1$ and $3-2$ bands involving lower levels with energies up to $\simeq 7500$~K formed in the stellar photosphere.
The HCN lines belong to bands $\nu_2(\pi)$, $2\nu_2(\sigma^+)$, $2\nu_2(\sigma^+)-\nu_2(\pi)$, $2\nu_2(\delta)-\nu_2(\pi)$, $3\nu_2(\pi)-\nu_2(\pi)$, $3\nu_2(\pi)-2\nu_2(\sigma^+)$, $3\nu_2(\pi)-2\nu_2(\delta)$, $3\nu_2(\phi)-2\nu_2(\delta)$, $4\nu_2(\sigma^+)-2\nu_2(\sigma^+)$, and $4\nu_2(\delta)-2\nu_2(\delta)$, and involve lower levels with energies up to $\simeq 3900$~K.
The apparently different vibrational excitation of CS and HCN might be an optical thickness effect related to the significantly higher Einstein coefficients of the CS lines in the observed spectral range ($A_{ul}\lesssim 40$~s$^{-1}$) compared to those of the HCN lines \citep[$A_{ul}\lesssim 4$~s$^{-1}$, see the Exomol Database;][]{exomol}, but other explanations such as the existence of a low HCN abundance gap near the stellar photosphere cannot be ruled out.
We have also identified \acet{} lines of different bands up to $\nu_4+3\nu_5(\delta_u)-2\nu_5(\delta_g)$, which involve lower levels with energies $\lesssim 2900$~K.
No SiS lines were detected even though its spectrum is partially covered by our observations.
There are still several unidentified features.
Some of them are probably lines of higher excitation hot bands of CS, HCN, and C$_2$H$_2$ but some features could be lines of other molecules.

The width at the baseline of the HCN ro-vibrational lines of the fundamental band $\nu_2(\pi)$ (at $\simeq 715$~\cm) is $\simeq 0.06-0.08$~\cm, about twice as high as for lines produced by gas expanding at $7.0-9.0$~\kms{} \citep[e.g.,][]{olofsson_1993a,izumiura_1995,schoier_2000,ramstedt_2014,massalkhi_2018b}, which would be of $\simeq 0.04$~\cm{} considering the resolving power of the spectrograph.
This implies maximum gas expansion velocities $\gtrsim 13$~\kms{} or local line widths at least as high as $\simeq 10$~\kms{} (see Section~\ref{sec:discussion.turbulence}).

\begin{figure}
  \centering
  \includegraphics[width=0.475\textwidth]{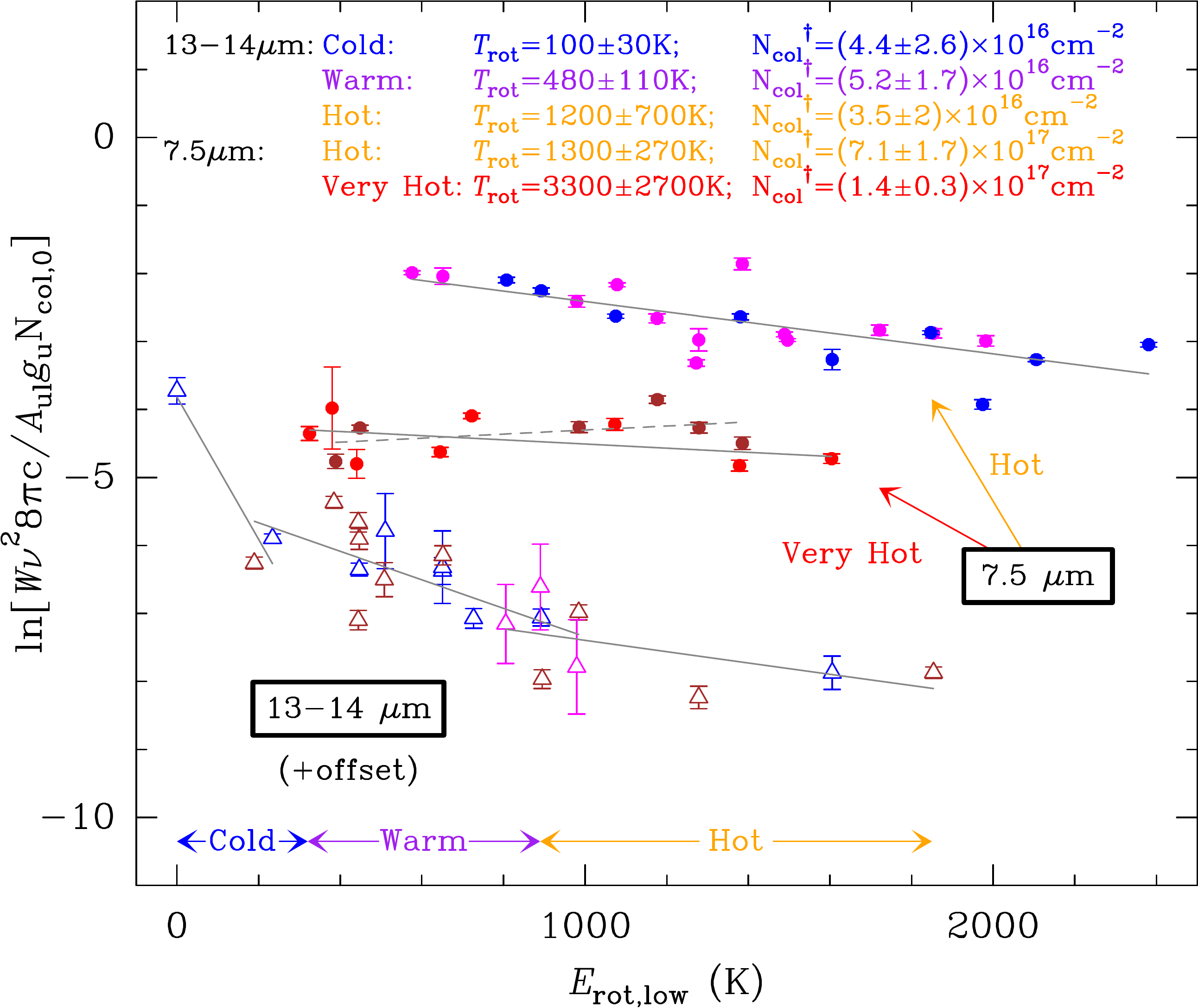}
  \caption{Ro-vibrational diagrams of absorption components of the infrared HCN lines at 7.5 and $13-14$~$\mu$m.
    The filled blue, magenta, brown and red dots correspond to the lines of the bands $2\nu_2(\sigma^+)$, $3\nu_2(\pi)-\nu_2(\pi)$, $4\nu_2(\sigma^+)-2\nu_2(\sigma^+)$, and $4\nu_2(\delta)-2\nu_2(\delta)$, respectively (7.5~$\mu$m).
    The empty blue, magenta, and brown empty triangles represent the lines of the bands $\nu_2(\pi)$, $2\nu_2(\sigma^+)-\nu_2(\pi)$, and $2\nu_2(\delta)-\nu_2(\pi)$ at $13-14~\mu$m.
    The dots related to the hot bands (except $4\nu_2(\sigma^+)-2\nu_2(\sigma^+)$ and $4\nu_2(\delta)-2\nu_2(\delta)$; see text) have been vertically shifted to make them coincide with the fundamental band at $13-14~\mu$m and with the first overtone at $7.5~\mu$m to get more populated data sets.
    An additional offset of $-2$ units has been added to the data at $13-14~\mu$m to improve visibility.
    Four populations with various rotational temperatures derived from the fitted solid straight lines can be distinguish (cold, warm, hot, and very hot).
    The dashed straight line is a fit to the lines of the $4\nu_2(\sigma^+)-2\nu_2(\sigma^+)$ band, which implies a negative rotational temperature (see text).
    $N_\subscript{col}^\dagger$ is defined as $N_\subscript{col,0}Z_\subscript{rot}e^{y_0}$ (Eq.~\ref{eq:y-intercept}).}
  \label{fig:f5}
\end{figure}

As shown above, the continuum of \ycvn{} is dominated by the central star at 7.5 and $13-14$~$\mu$m and we can assume that the absorption components of all the ro-vibrational lines (Figures~\ref{fig:f1} and \ref{fig:f2}) are constrained to be produced in a region as small as $\simeq 0\farcs014$ ($\simeq 2\rstar$) to first approximation.
With this in mind, we have done a ro-vibrational diagram for the absorption components of the HCN lines that can be found in Figure~\ref{fig:f5}.

The fits to the absorption components have been done by adopting the function $e^{-\tau_\nu}$, where $\tau_\nu=\tau_0e^{-(\nu-\nu_0)^2/\sigma^2}/\left(\sigma\sqrt{\pi}\right)$ and $\sigma=\textnormal{FWHM}/\left(2\sqrt{\ln{2}}\right)$.
In the fits to the P-Cygni profiles we have added the function $K\left(1-e^{-\tau^\prime_\nu}\right)$ to fit the emission components at the same time as the absorptions (see Appendix~\ref{sec:rovib.diagram}).
The $\tau_\nu$ and $\tau^\prime_\nu$ are different and involve also different Doppler shifts.

The total data set plotted in Figure~\ref{fig:f5} contains the opacities of the more isolated, unperturbed lines of several different bands: $2\nu_2(\sigma^+)$, $3\nu_2(\pi)-\nu_2(\pi)$, $4\nu_2(\sigma^+)-2\nu_2(\sigma^+)$, and $4\nu_2(\delta)-2\nu_2(\delta)$ at 7.5~$\mu$m, and $\nu_2(\pi)$, $2\nu_2(\sigma^+)-\nu_2(\pi)$, and $2\nu_2(\delta)-\nu_2(\pi)$ at $13-14~\mu$m.
All the ro-vibrational bands at 7.5 and $13-14~\mu$m separately are compatible between them except $4\nu_2(\sigma^+)-2\nu_2(\sigma^+)$, which shows a growing dependence with the increasing rotational energy and thus a negative rotational temperature.
We suspect that these lines are affected by either radiation transfer mechanisms not considered in our analysis or errors in the spectroscopic data (see Section~\ref{sec:discussion.temperatures}).
Consequently, we have not used the lines of this band in the fits.

The ro-vibrational diagram discloses the existence of four different HCN populations, cold, warm, hot, and very hot, with rotational temperatures of $100\pm 30$, $480\pm 110$, $1300\pm 270$, and $3300\pm 2700$~K.
The poor accuracy of the temperature of the very hot population is due to the the high dispersion inherited from lines with a poorer S/N.
The Q branches, even though they are almost completely covered by our observations, are ineffective to study cold gas in the outer envelope due to the typical width of the lines, which is significantly larger than the line separation as the rotational number decreases.

The strong vibrational excitation undergone by HCN implies a significant population in at least four vibrational states (from the ground state up to the $3\nu_2$ state) and four different vibrational temperatures.
These temperatures, which are unknown a priori and hard to determine for most of the states from the ro-vibrational diagram due to the data dispersion, make it difficult to estimate the vibrational partition function of HCN, $Z_\subscript{vib}$.
Therefore, we can only derive lower and upper limits for the column density.

The lower limits are $(4.4\pm 2.6)\times 10^{16}$, $(5.2\pm 1.7)\times 10^{16}$, $(3.5\pm 1.0)\times 10^{17}$, and $(1.4\pm 0.3)\times 10^{17}$~\cmm{} for the cold, warm, hot, and very hot populations, respectively.
The total column density is thus $N_\subscript{col}\gtrsim 4.1\times 10^{17}$~\cmm.

We have assumed that the maximum vibrational partition function would be reached for HCN vibrationally excited at LTE.
The upper limits for $Z_\subscript{vib}$ are thus $\simeq 1.00$ at 100~K, $\simeq 1.28$ at 480~K, $\simeq 3.76$ at 1300~K, and $\simeq 19.0$ at 3300~K.
These calculations mean that $N_\subscript{col}\lesssim 5.1\times 10^{18}$~\cmm.
Hence, the HCN column density in \ycvn{} is $0.026-0.32$ times as high as that found in \irc{} \citep*[$\simeq 1.6\times 10^{19}$~\cmm;][]{fonfria_2008}, which is quite high even for the lower limit considering that \ycvn{} displays a mass-loss rate about 190 times lower than \irc{} \citep*[$\simeq 1.4\times 10^{-7}$ and $2.7\times 10^{-5}$~\mlr, respectively;][]{ramstedt_2014,guelin_2018}.

The mid-IR lines trace the gas typically up to a few tens of stellar radii.
Outwards, the bulk of the HCN is in the vibrational ground state and the mid-IR lines are no longer sensitive to the physical conditions in these outer regions.
However, we can explore the outer envelope with the pure rotational lines observed in the millimeter range.
These observations comprise two lines in the vibrational ground state ($J=1-0$ and $3-2$) and three $J=3-2$ lines in the first vibrationally excited state, $\nu_2$, with different parities $e-f$ and vibrational angular momentum ($\ell=0,~\pm 1$, and $\pm 2$ for the $\sigma^+$, $\pi$, and $\delta$ states, respectively).

The $v=0$ $J=1-0$ lines of HCN and H$^{13}$CN show a noticeable hyperfine structure involving three components contained within an interval of $\simeq 15$~\kms{} (Figure~\ref{fig:f3}).
\citet{izumiura_1987,izumiura_1995} and \citet{dinh-v-trung_2000} showed that large fractions of these lines are maser emission.
Therefore, finding the thermal emission might be difficult and any result derived from this line has to be taken with care.

The $v=0$ $J=3-2$ HCN and H$^{13}$CN lines are mostly parabolic, which means that they are optically thick and the emitting region is unresolved.
They display a similar intensity, which can be explained by the low isotopic ratio $^{12}$C/$^{13}$C~$\simeq 3$ for \ycvn{} \citep{abia_2017}.
These lines also show a hyperfine structure with six components separated by a maximum velocity of $\simeq 4$~\kms.
They show several interesting features: (1) their peak emissions are located at $\simeq +2$~\kms{} from the systemic velocity, adopting the frequencies without hyperfine structure in Table~\ref{tab:table3} for the lines, (2) the lines are narrower by $2-3$~\kms{} than they should be for a terminal expansion velocity of $\simeq 8$~\kms, (3) they have weak, blue-shifted contributions around $-8$~\kms{} spanning $\simeq 3$~\kms, and (4) they have red-shifted wings.
All these features can be simultaneously explained a priori if: (a) the line profiles are affected by the hyperfine structure and/or there is gas expanding at velocities higher than the terminal velocity, and (b) a blue-shifted fraction of the lines are strongly self-absorbed.

The rotational HCN lines in the $\nu_2(\pi)$ vibrational state with a different parity \textit{e-f} display an unexpectedly high intensity ratio of one order of magnitude, where the $e$ line is stronger.
Their $A$-Einstein coefficients are very similar ($7.422\times 10^{-4}$~s$^{-1}$ for the $e$ line and $7.535\times 10^{-4}$~s$^{-1}$ for the $f$ line; \citealt{cernicharo_2012}) and the energies of their lower levels are essentially the same.
These lines display very similar triangular profiles.
However, the $e$ line exhibits a moderately strong, narrow red-shifted peak at $\simeq +3$~\kms{} from the systemic velocity (Figure~\ref{fig:f3}).
The only way to explain these two remarkable differences is that maser emission takes part in the HCN $\nu_2^{1e}$ $J=3-2$ line.
The corresponding H$^{13}$CN lines are much more similar and significantly weaker than the lines in the vibrational ground state, which suggests that they involve rotational levels that are thermally populated.

We note that while the intensities of the $J=3-2$ lines of HCN and H$^{13}$CN in the ground and in the $2\nu_2(\sigma^+)$ vibrational states are of the same order of magnitude ($\sim 1$ and 0.01~K, for each vibrational state) there are differences of two and one orders of magnitude for the lines $\nu_2^{1e}$ $J=3-2$ and $\nu_2^{1f}$ $J=3-2$, respectively.
This may imply that the HCN line $\nu_2^{1f}$ $J=3-2$ is also showing maser emission though much weaker than $\nu_2^{1e}$ $J=3-2$.

\begin{figure*}
  \centering
  \includegraphics[width=0.9\textwidth]{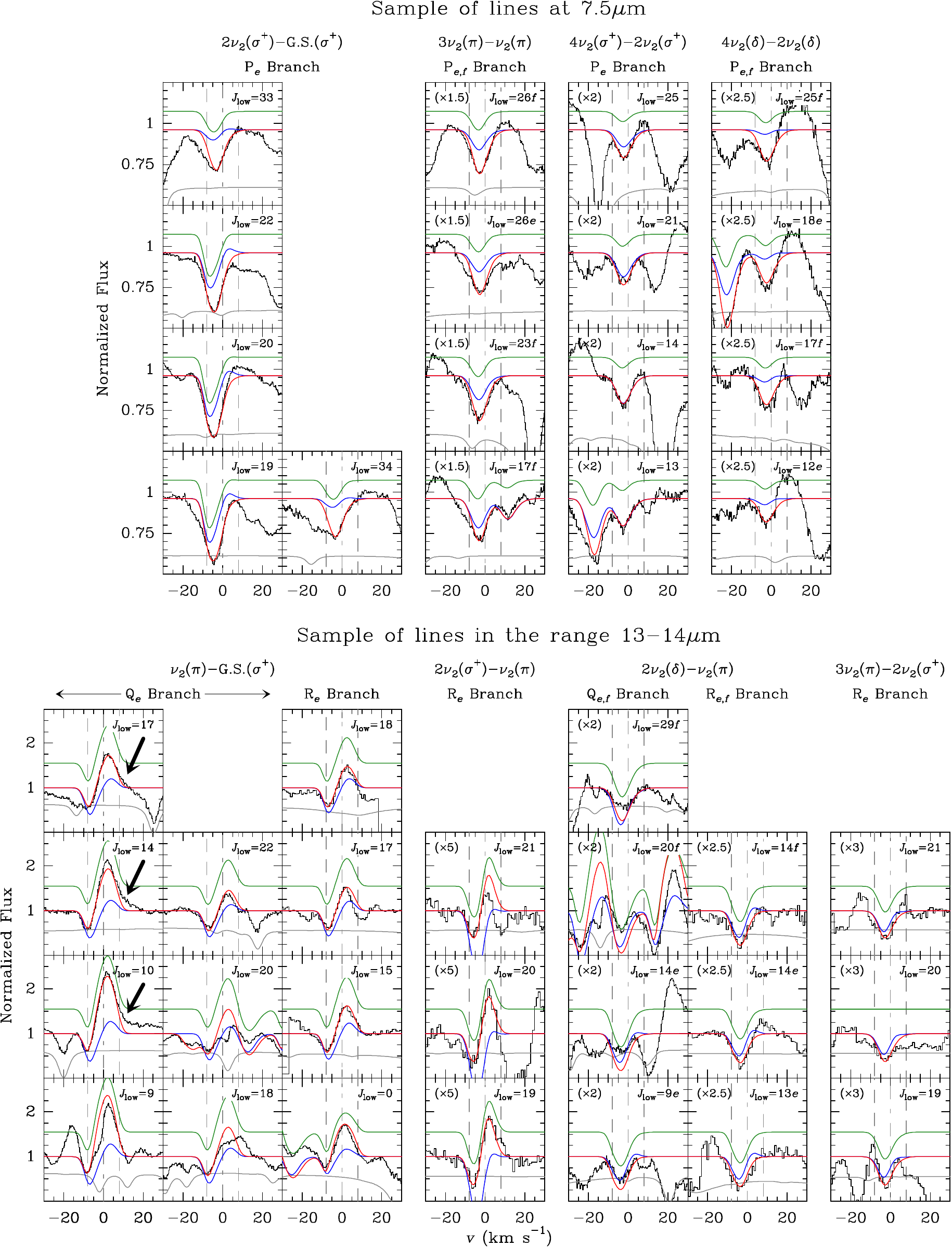}
  \caption{Sample of observed HCN lines in the mid-IR (black histogram) with their best fits (spherically symmetric models; red curves) grouped in bands.
    The blue solid lines in the range $13-14~\mu$m have been calculated with the spherically symmetric envelope model that better fits the lines at $7.5~\mu$m and vice versa (see Table~\ref{tab:table2} and text).
    The green solid lines have been calculated with the asymmetric envelope model (see Section~\ref{sec:asymmetric.model}).
    They are vertically shifted to improve the visibility.
    The gray curves are the atmospheric transmission.
    The black arrows point at possible red-shifted wings.
    The vertical dashed-dotted and dashed lines are the systemic velocity (0~\kms) and the terminal Doppler velocities derived in this work ($\simeq 8$~\kms), respectively.
    All the fits are reasonably good except those to the lines of band $2\nu_2(\delta)-\nu_2(\pi)$, Q branch (see text).
}
  \label{fig:f6}
\end{figure*}

\subsection{The best fit to the observed HCN lines}
\label{sec:bestfit}

\subsubsection{Spherically symmetric model}
\label{sec:symmetric.models}

\begin{figure}
  \centering
  \includegraphics[width=0.475\textwidth]{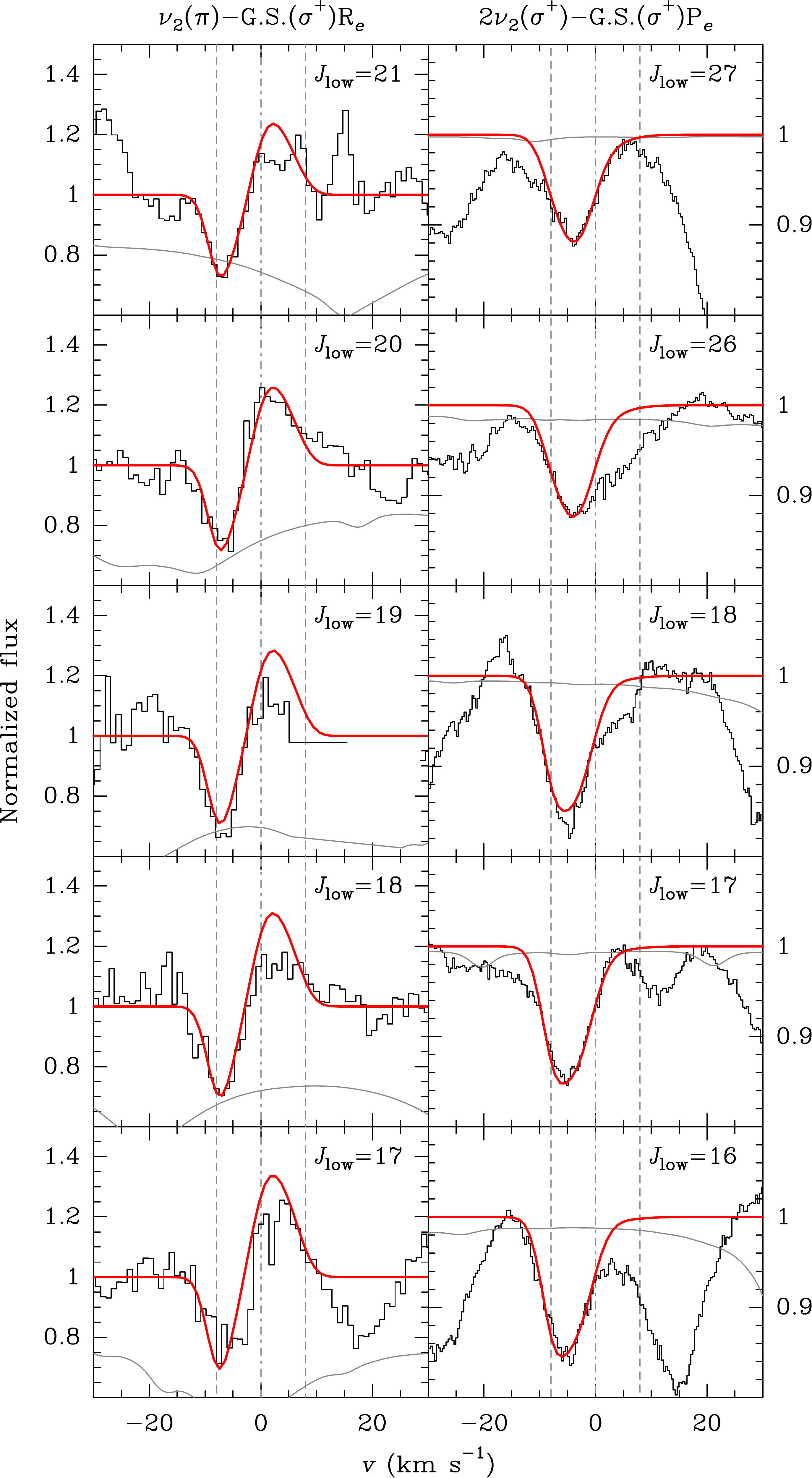}
  \caption{Sample of observed H$^{13}$CN lines in the mid-IR (black histogram) with their best fits (spherically symmetric models; red curves) grouped in bands.
    The gray curves are the atmospheric transmission.
    The vertical dashed-dotted and dashed lines are the systemic velocity (0~\kms) and the terminal Doppler velocities derived in this work ($\simeq 8$~\kms), respectively.
}
  \label{fig:f7}
\end{figure}

The fits to some of the lines obtained using the radiative transfer code can be found in Figure~\ref{fig:f6} and the derived parameters are included in Table~\ref{tab:table2}.
The H$^{13}$CN lines have been fitted by varying only the abundance and the photodissociation radius as no strong excitation temperature variations are expected (Figure~\ref{fig:f7}).

\begin{figure}
  \centering
  \includegraphics[width=0.475\textwidth]{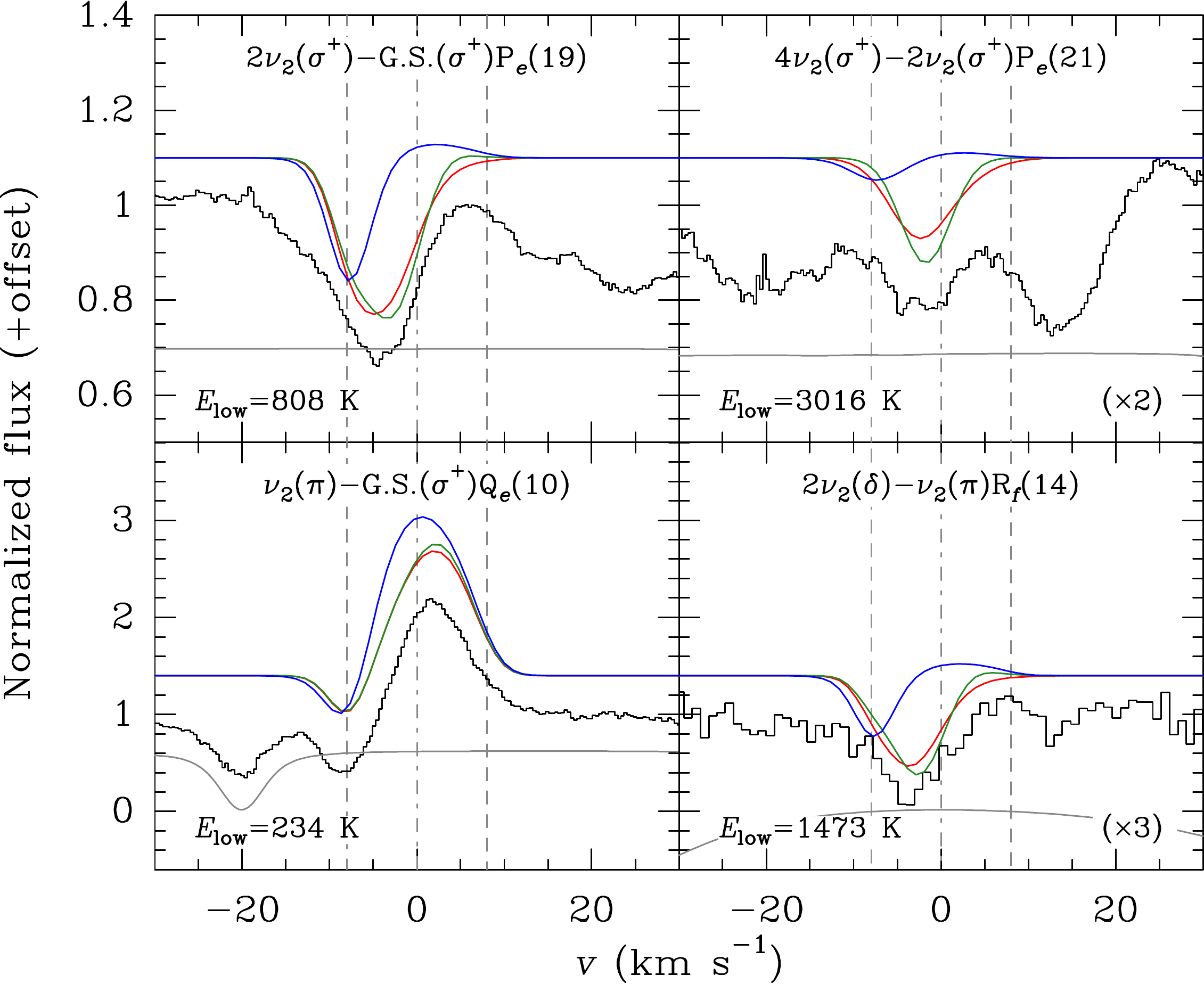}
  \caption{The effect of varying the turbulent velocity and the gas expansion velocity at the stellar surface (green and blue, respectively) on the synthetic profiles coming from different regions of the envelope.
      The observations are the black histograms and the best fits are plotted in red.
      In the green curves, the turbulent velocity has been set to 5~\kms.
      The blue curves have been calculated with a constant gas expansion velocity of 8~\kms{} throughout the whole envelope.
      These curves have been vertically shifted for clarity.
      $E_\subscript{low}$ is the energy of the lower level of the ro-vibrational transitions.
      The gray solid curves are the atmospheric transmission.
      The vertical gray dotted-dashed and dashed lines indicate the systemic velocity and the Doppler shifts related to the terminal velocity.}
  \label{fig:f8}
\end{figure}

An initial assumption of a constant radial velocity profile equal to 8~\kms{} from the stellar photosphere and outwards \citep{izumiura_1995,ramstedt_2014}, produces synthetic profiles for the 7.5~$\mu$m region narrower than those observed, considering the EXES' spectral resolution (full width at half depth $\simeq 2.5$~\kms{} compared to $\simeq 5.4$~\kms{} for low-$J$ lines of the overtone $2\nu_2(\sigma^+)$).
Additionally, their absorption components peak at Doppler velocities with respect to the systemic one $\gtrsim -7.5$~\kms, substantially different from the observed peak velocities ($\gtrsim -4.5$~\kms).
This disagreement can be resolved with a constant gas velocity gradient close to the star, which represents the acceleration of the ejected gas, followed by a constant expansion velocity of 8~\kms{} (Figure~\ref{fig:f8}, Table~\ref{tab:table2}).

The fit to all the ro-vibrational lines indicates that the width at the baseline of the HCN lines of the fundamental band can be explained by strong turbulence in the vicinity of \ycvn.
However, this model might be too simple as a few HCN lines at 14.0~$\mu$m show unexpected wings in the emission component that cannot be accounted neither with the terminal expansion velocity nor with local line widths due to turbulence (Figures~\ref{fig:f6} and \ref{fig:f8}).
Furthermore, none of these HCN lines are overlapped with sufficiently strong acetylene features.

The optically thick pure rotational lines $v=0$ $J=3-2$ of HCN and H$^{13}$CN also show red-shifted high velocity wings (Figure~\ref{fig:f3}).
Our current spherically symmetric envelope model cannot reproduce these wings, even considering the hyperfine structure, which broadens the lines at the baseline.
We give a possible solution to this problem in Section~\ref{sec:asymmetric.model}.

The best fit to the HCN lines at 7.5~$\mu$m is compatible with a constant profile for the abundance with respect to H$_2$ in the gas acceleration region ($r\simeq 1-30\rstar$).
However, the vibrationally excited lines at $13-14~\mu$m cannot be fitted with such an abundance distribution and, instead, an abundance at the stellar photosphere lower by a factor of $\simeq 4$ is needed.
The lines at 7.5~$\mu$m indicate that HCN is in rotational LTE close to the photosphere but \trot{} is steeper than \tk{} moving outwards.
The $13-14~\mu$m lines suggest that HCN may be out of rotational LTE even close to the stellar photosphere (see Figure~\ref{fig:f6}).

\begin{figure}
  \centering
  \includegraphics[width=0.475\textwidth]{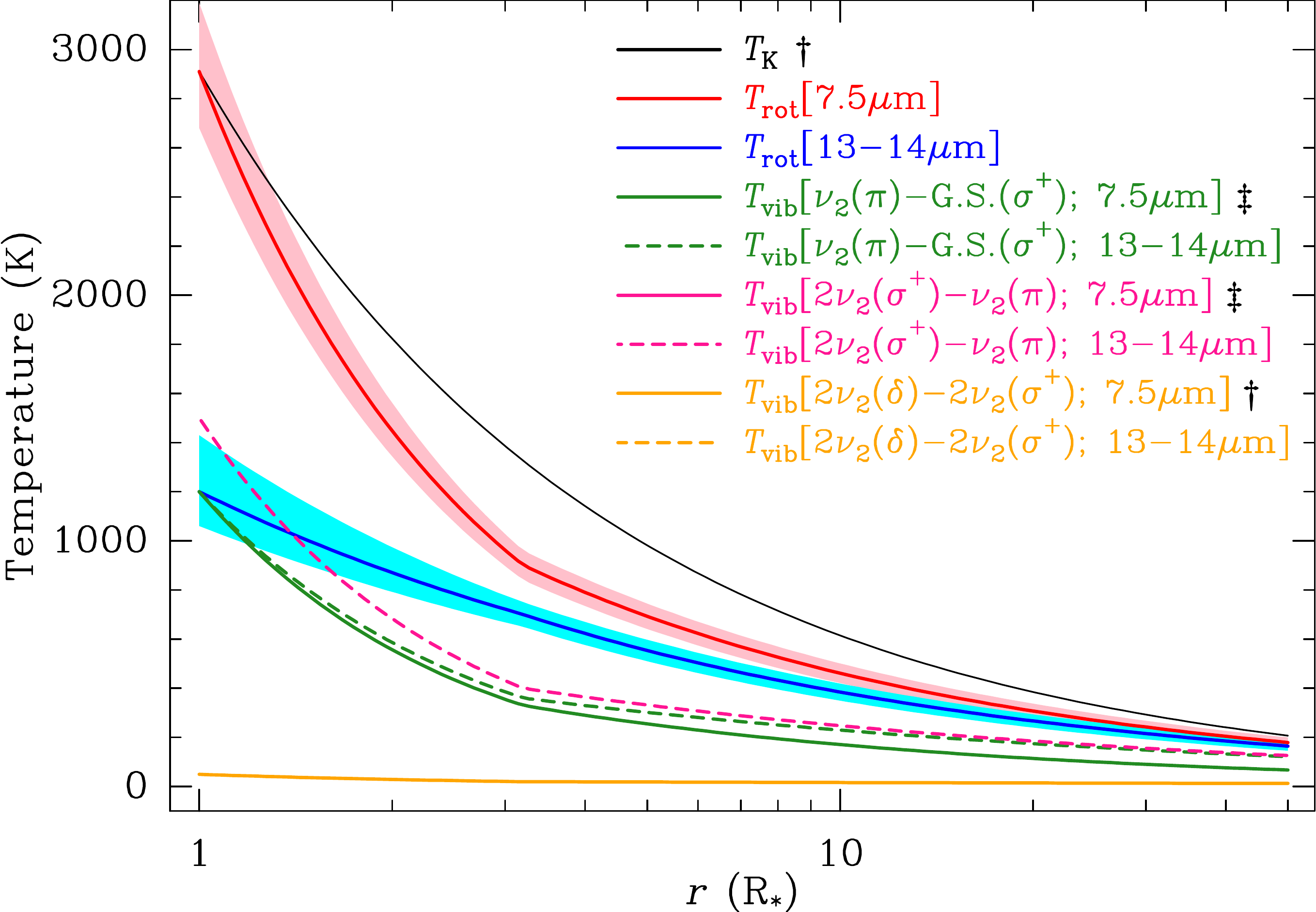}
  \caption{Temperatures involved in the envelope models.
      The kinetic temperature, \tk, has been adopted from \citet{guelin_2018}.
      The rotational and vibrational temperatures, \trot{} and \tvib, have been derived in this work (Table~\ref{tab:table2}).
      The shaded regions are the $1\sigma$ errors.
      The uncertainties of the vibrational temperatures have not been plotted for sake of clarity.
      The temperatures with the same symbol ($\dagger$ or $\ddagger$) coincide.}
  \label{fig:f9}
\end{figure}

Cold and warm gas is strongly involved in the formation of the pure rotational lines $v=0$ $J=3-2$ of HCN and H$^{13}$CN.
These lines cannot be accurately reproduced, which means that our model is also too simple for the outer envelope.
However, our synthetic lines partially fit the profiles of the observed data.
Their intensities are compatible with a constant abundance with respect to H$_2$ of $\simeq 1.3\times 10^{-4}$ and an isotopic ratio $^{12}$C/$^{13}$C~$\simeq 2.5$ throughout the whole envelope, which is valid also for the mid-IR spectra.

The red-shifted side of the $v=0$ $J=3-2$ rotational HCN line can be fitted assuming that the abundance and excitation temperatures derived from the ro-vibrational lines in the mid-IR hold throughout the whole CSE, and if the HCN photodissociation shell is located at $\simeq 170\rstar$ ($\simeq 1\farcs2$) from the star.
The rotational temperature at this distance from the star drops down to $\simeq 85-90$~K.
Our envelope model cannot describe the blue-shifted part of this line as the self-absorption mentioned above is not strong enough.
The results derived for H$^{13}$CN are compatible with a photodissociation region placed about 15\rstar{} closer to the star (Table~\ref{tab:table2}).

The fit to the $v=0$ $J=1-0$ HCN line requires maser emission to reproduce the observed central peak in addition to the red-shifted narrow spike that the line shows at $\simeq +4$~\kms.
Thermal emission is only noticeable in the wings of this line and the rest is dominated by maser emission \citep{dinh-v-trung_2000}.
The H$^{13}$CN $v=0$ $J=1-0$ line is very similar and most of it has to be explained by maser emission as well.

In summary, we have not found a spherically symmetric envelope model able to simultaneously reproduce all the observations ($7.5~\mu$m, $13-14~\mu$m, and mm range -- see Figure~\ref{fig:f6}).
Thus, we propose an asymmetric envelope model in the next Section that roughly describes most of the observed lines.

\subsubsection{Asymmetric model}
\label{sec:asymmetric.model}

A multidimensional approach involving at least 2D asymmetries seems necessary to find a more realistic envelope model that explains all the features simultaneously.
Nevertheless, this model is a rough approximation and aims to provide some qualitative clues about the actual CSE of \ycvn.

\begin{figure*}
  \centering
  \includegraphics[width=\textwidth]{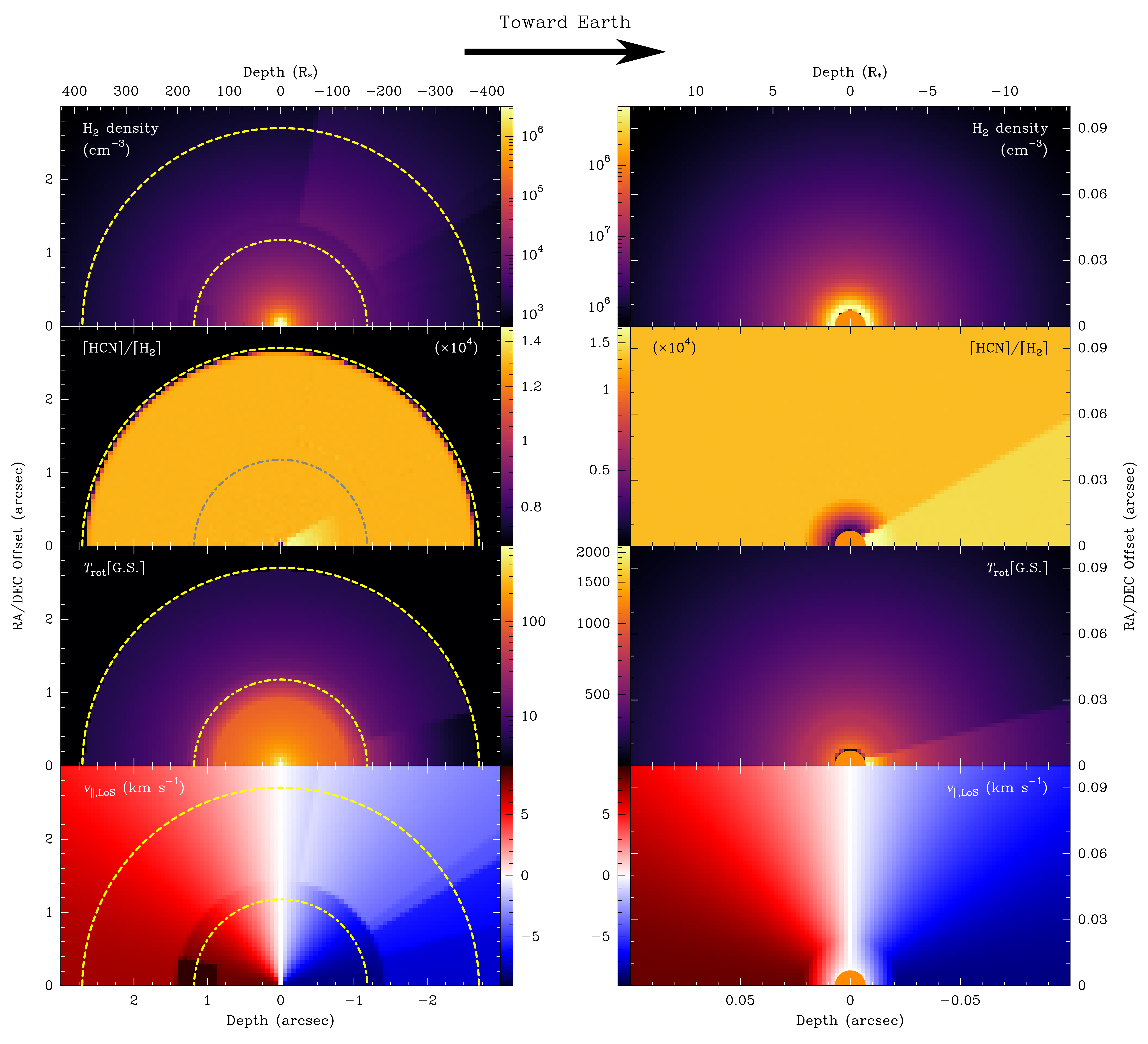}
  \caption{H$_2$ density, HCN abundance with respect to H$_2$, rotational temperature in the vibrational ground state, and gas expansion velocity projected onto the line-of-sight derived from the asymmetric envelope model.
    The abscissas are the depth across the envelope measured from the closer position to Earth backwards.
    The ordinates are the position in the plane of the sky along an undetermined direction since the 2D envelope model shows cylindrical symmetry around the line-of-sight.
    Each quantity has been plotted twice, at large scale (left) and at small scale (right).
    In the maps to the left, the yellow dotted-dashed and dashed circles represent the HCN dissociation radii derived in this work for the spherically symmetric envelope model ($\rph=168\rstar$) and for the model that accounts for the self-absorption in the pure rotational lines ($\rph=380\rstar$), respectively (see text).
    In the maps to the right, the filled orange circle is the central star.
}
  \label{fig:f10}
\end{figure*}

Most of the inner envelope characteristics were reasonably well described by the spherically symmetric model derived from the analysis of the lines at $13-14~\mu$m.
The spherically symmetric model derived from the lines at $7.5~\mu$m, with slight modifications, was used to describe the conical region in front of the star with the star at its apex, its axis parallel to the line-of-sight, and an aperture of $30^\circ$.
The addition of shells with different physical and chemical conditions beyond the photodissociation radius for the spherically symmetric model have allowed us to describe the pure rotational lines, in particular the blue-shifted self-absorption components and the red-shifted wings of the $v=0$ $J=3-2$ lines.
Figure~\ref{fig:f10} contains maps of the H$_2$ density, the HCN abundance with respect to H$_2$, the rotational temperature of HCN, and the gas expansion velocity projected along the line-of-sight.

The wings observed in the $J=3-2$ rotational lines (Figure~\ref{fig:f3}) are reproduced after increasing the gas expansion velocity in the region $125-200\rstar$ to $\simeq 9-10$~\kms{} along the line-of-sight in the CSE hemisphere behind the star.
We stress that these wings cannot be explained by the hyperfine structure of HCN (and H$^{13}$CN).

The formation of the self-absorptions is mostly related to gas expanding at velocities as low as 4.5~\kms.
They can be roughly replicated by adding cold HCN in the region of the CSE corresponding to the hemisphere facing us that expands obliquely with an inclination with respect to the plane of the sky greater than $10^\circ$ beyond $\simeq 200\rstar$.
In this region of the envelope, the required rotational temperature ranges from 5 to 15~K.

The new photodissociation radius ($\simeq 380\rstar$), the uncertainty of which is as least as high as the line flux calibration errors ($\simeq 20\%$), suggests that there could be cold HCN in the outer envelope that do not significantly contribute to the emission of the $v=0$ $J=3-2$ lines as the bulk of them come from the region of the CSE inwards from $\simeq 170\rstar$.
The HCN column density in front of the star calculated with this model is $\simeq 3.5\times 10^{18}$~\cmm.

The comparison of the observed lines and those produced by our spherically symmetric and asymmetric models (red and blue in Figure~\ref{fig:f3}) confirms our assessment about which rotational lines show maser emission (Section~\ref{sec:molecular.emission}).
The emissions of the HCN and H$^{13}$CN lines $v=0$ $J=1-0$ as well as the HCN lines $\nu_2^{1e}$ $J=3-2$ and $\nu_2^{1f}$ $J=3-2$ are essentially of maser nature.
All these masers can be roughly reproduced by adding a few compact maser emitting regions preferentially located at the stellar photosphere and at the end of the acceleration region.
In the latter shell, the gas density and its expansion velocity are high enough to produce maser peaks at any Doppler velocity in the millimeter spectrum, depending on the expansion direction.
The maser contributions produced by our model are usually narrower than the observed ones but we think that they can be described if the maser emitting region is radially more extended than we have assumed, it covers a larger part of the gas acceleration zone, or it comprises different isolated emitting shells or arcs.

\section{Discussion}
\label{sec:discussion}

\subsection{The continuum emission}

\subsubsection{The effective temperature of \ycvn}
\label{sec:effective.temperature}

The effective temperature derived from our analysis is $2910\pm 70$~K.
This value is higher than the effective temperature of 2860~K reported by \citet{ohnaka_1999} and 2760~K proposed by \citet{bergeat_2001}, which was reinforced by \citet{debeck_2010}.
\citet{goebel_1980} determined that the effective temperature ranges from 2750 to 2900~K from a series of infrared observations taken over two years.
All these measures differ by 150~K or less suggesting that the disagreement is only apparent and it is likely produced by observational uncertainties and stellar pulsation effects.

However, our estimate is not compatible with the effective temperatures assumed or derived by others \citep*[e.g.,][]{teyssier_2006,schoier_2013,ramstedt_2014}, which are more than 25\% lower.
These results are based on photometric observations in the J ($1.25~\mu$m) to M ($4.8~\mu$m) bands.
These measurements are going to be possibly confused by effect of the strong molecular bands of CO, C$_2$, C$_3$, C$_2$H$_2$, CN, and HCN that exist below $5~\mu$m.

\subsubsection{Dust grains composition and distribution throughout the CSE}
\label{sec:discussion.dust}

One of the most interesting results of our fit is the absence of amorphous carbon close to the star, contrary to what occurs in \irc.
In the latter, the bulk of the carbonaceous material condenses onto the dust grains at $\simeq 5\rstar$ from the center of the star \citep{fonfria_2008}, where the gas kinetic temperature is $\simeq 800-1000$~K.
This condensation temperature is reached in \ycvn{} in the shell that extends from $4.9\rstar$ to $6.8\rstar$, where the gas density is $\simeq 1-2\times 10^7$~cm$^{-3}$.
However, the \textit{ISO} observations indicate that most of the AC emission comes from a region of the CSE where the gas kinetic temperature is below 100~K.
In this region the gas is rarefied (gas density $\sim 10^4$~cm$^{-3}$, typical of dense molecular clouds), highly decreasing the depletion rate of any possible refractory molecular species that could produce the AC in the grains and, thus, increasing the grain growth timescale beyond $\sim 1$~Myr \citep{hirashita_2000}.
This timescale is too long compared to the typical kinetic times involved in the evolution of AGB CSEs ($\sim 100-1,000$~yr).
The low kinetic temperature comes with an increase in the density of dust grains by a factor of 15 in order to reproduce the observed flux densities, which can also be explained as an increase of the grain size by a factor of $\simeq 2.5$, if we assume optically thin grains.

This result can be explained if the condensation of the refractory carbon-bearing molecules onto the SiC dust grains which form AC close to the star stopped in the past and the AC we are seeing is the remnant of this process.
It would have ended about $200-300$~yr ago, if we assume an expansion velocity for the dust grains twice as high as the terminal gas expansion velocity of $\simeq 8$~\kms.
A dusty shell like this, which we place at $\simeq 1\farcs5$ from the star, should not be confused with the detached shell found by \citet{young_1993} and \citet{izumiura_1996}, which is located at $\simeq 180\arcsec-190\arcsec$ from the star.
Thus, we could consider that these kind of dusty shells might be formed periodically or, at least, that the latter was not formed by an unique event.

\subsection{Considerations of the molecular envelope model}
\label{sec:discussion.hcn}

\subsubsection{The gas expansion velocity}
\label{sec:discussion.velocity}

The gas expansion velocity profile derived during our analysis is in good agreement with a typical profile produced by gas acceleration by radiation pressure on dust grains \citep*[e.g.,][]{goldreich_1976,bowen_1988,decin_2006}.
The gas accelerates from a low velocity roughly at the photosphere over a zone of a few stellar radii to reach a terminal velocity similar to previously reported values derived from pure rotational lines \citep{izumiura_1995,ramstedt_2014,massalkhi_2018b}.
The condensation of SiC seems to be responsible for the gas acceleration.

However, this gas expansion velocity field cannot satisfactorily explain the red-shifted wings found in the Q branch mid-IR lines of the HCN band $\nu_2(\pi)$ (see Section~\ref{sec:symmetric.models} and Figure~\ref{fig:f6}).
These wings are not present in high-$J$ lines and they are not expected to be noticeable in many other lines of this band due to the high line density of the observed spectral ranges.
They could mean that the gas responsible, which would be located at a few tens of \rstar, is expanding faster than the terminal velocity.

Moreover, the $J=3-2$ lines in the vibrational ground state of HCN and H$^{13}$CN display evidence of a higher gas expansion velocity (Figure~\ref{fig:f3}).
These lines are optically thick so that only an incomplete shell with gas expanding faster than the terminal velocity and located in the outer envelope can explain this effect.
We note that the need for an incomplete shell cannot be a consequence of an exceedingly narrow telescope main beam since the HCN emitting region of the envelope is completely covered by it ($2\rph(\textnormal{HCN})\simeq 380\rstar\simeq 5\farcs3$, HPBW$(266~\textnormal{GHz})\simeq 9\farcs2$).
Thus, both the IR and the mm lines could be tracing gas placed between a few tens of stellar radii and $\simeq 200\rstar$, expanding at up to $\simeq 10$~\kms.
This velocity could be higher if the gas expands along a direction different from the line-of-sight.

\subsubsection{The turbulent velocity}
\label{sec:discussion.turbulence}

The fits to the mid-IR lines suggest that close to the photosphere the local line width of the HCN lines is $\simeq 10$~\kms.
The thermal broadening for HCN, at the kinetic temperature prevailing near the photosphere of \ycvn, is $\simeq 1$~\kms.
Considering a Maxwell-Boltzmann velocity distribution, the gas turbulent velocity is therefore as high as $\simeq 6$~\kms.

For \irc, many lines coming from the vicinity of the photosphere, which have a line width ranging from 5 to 8~\kms, have been found so far \citep{patel_2009,cernicharo_2011,velilla-prieto_2015}.
These line widths represent turbulent velocities of $\simeq 3-5$~\kms, which are considered to be related to strong turbulence typical of the innermost layers of the envelope of an AGB star, where efficient matter ejection mechanisms are enabled.

Hence, the comparison between the turbulent velocity for \ycvn{} and \irc{} suggests that (1) unexpectedly violent mechanisms at work close to the photosphere of \ycvn{} produce very strong turbulence, (2) the turbulent velocity close to the photosphere of \irc{} is underestimated, or (3) the line width of the HCN lines observed in \ycvn{} is produced by turbulence along with other gas acceleration mechanisms not included in our model.
This latter scenario is based on the fact that the lack of spatial resolution could lead us to confuse turbulence in the circumstellar envelope with other photospheric phenomena such as higher velocity gas ejections, convective cells or the pulsating movement of the photosphere \citep{freytag_2017}.
The high Doppler shifts and FWHMs of atomic lines that have been observed in red supergiants and Miras for some time \citep[e.g.,][]{smith_1989,hinkle_2002,gray_2008,stothers_2010} would support the existence of these phenomena.

\subsubsection{HCN abundance with respect to H$_2$ and column density}
\label{sec:discussion.abundance}

The fits to all the data are compatible with an HCN abundance of $\simeq 1.3\times 10^{-4}$ throughout most of the envelope with an anomaly very close to the photosphere that implies a decay of the abundance down to $3.5\times 10^{-5}$.
The error for this quantity in the inner envelope is better than 10\% due to the redundancy provided by the high number of observed vibrationally excited lines that come from the vicinity of the star.
The impossibility of determining the thermal emission of the HCN $v=0$ $J=1-0$ line and the calibration errors of the pure rotational lines lead us to think that a factor of 2 for the abundance in the outer envelope could be a better estimate than the one based on the sensitivity of our model to the free parameters ($\simeq 15\%$; Table~\ref{tab:table2}).
An additional uncertainty related to the determination of the distance to \ycvn{} has to be considered.
In the current work we have adopted the distance published in the Gaia EDR3 \citep[$310\pm 17$~pc;][]{gaia_edr3} but shorter distances were proposed before [$\simeq 220$~pc \citep{abia_2000,schoier_2013} and $232\pm 13$~pc \citep[Gaia Data Release 2;][]{gaia_dr2}].
Consequently, the HCN and H$^{13}$CN abundances might be $30$\% lower than we propose.

This abundance profile is based on a gas density $\propto r^{-2}$ beyond the photosphere but this profile could be steeper in the region between the photosphere and $r\simeq 2-3\rstar$ \citep*[e.g.,][and references therein]{agundez_2020}.
The adoption of a gas density profile like this would reduce the HCN abundance close to the star with respect to that over the rest of the CSE even more as the HCN column density has to be the same to reproduce the observations.
Therefore, the abundance for this region of the CSE derived in this work has to be taken as an upper limit.

The average column density for HCN that we have derived from our fits (see Table~\ref{tab:table2}) is $(2.1\pm 1.1)\times 10^{18}$~\cmm{} and we proposed in Section~\ref{sec:molecular.emission} an upper limit of $\simeq 5.1\times 10^{18}$~\cmm{} under vibrational LTE.
However, the detailed analysis of the lines has made clear that HCN is always vibrationally out of LTE (see Section~\ref{sec:bestfit} and Figure~\ref{fig:f9}).
More precisely, the vibrational partition function derived from our best fits to the mid-IR lines is below $\simeq 4.2$ (it is below $\simeq 19$ under vibrational LTE; Section~\ref{sec:bestfit}).
Thus, the HCN column density derived from the ro-vibrational diagram considering the vibrational temperatures obtained from the modeling process is $(1.2\pm 0.3)\times 10^{18}$~\cmm, fully compatible with the result of our fits.

An observational estimate of the HCN abundance in the inner layers of the envelope of \ycvn{} has not been reported to date (as far as we know in the literature) but the photospheric abundance was determined to be $1.9\times 10^{-5}$ by \citet{olofsson_1993b}.
This abundance is the value in the upper layers of a static LTE model atmosphere which reproduced the results of \citet{lambert_1986}.
It is compatible with the value that we have derived from the analysis of the observations in the $13-14~\mu$m range but it is 7 times lower than the one inferred from the spectrum at 7.5~$\mu$m.
This disagreement can be explained by asymmetries or clumpiness around the star, which would reduce the HCN column density without diminishing its abundance, or by the gas-phase chemistry in the vicinity of the star, where the HCN abundance could be significantly lower than in the rest of the envelope \citep[e.g.,][]{fonfria_2008,agundez_2020}.

The rest of the abundance estimates in the literature have been derived apparently from the analysis of the $v=0$ HCN lines $J=1-0$, $3-2$, and $4-3$.
\citet{olofsson_1993b} reported an HCN abundance with respect to H$_2$ that follows a Gaussian profile with a value at the stellar photosphere of $8.3\times 10^{-4}$ and a photodissociation radius of $1.2\times 10^{15}$~cm ($\simeq 50\rstar$ for us).
The average abundance is $\simeq 7.3\times 10^{-4}$, i.e., about 6 times higher than ours.
This disagreement can be due to three reasons:
(1) they used only the HCN $v=0$ $J=1-0$, which shows maser emission \citep[e.g.,][]{dinh-v-trung_2000}, 
(2) their photodissociation radius is very small compared with ours ($\simeq 170-380\rstar$),
and (3) the different distance from the star and mass-loss rate that these authors used \citep[$D\simeq 290$~pc and $\dot M\simeq 4-6\times 10^{-8}$~\mlr;][]{olofsson_1993a}.

\citet{bachiller_1997} gave an average CN abundance with respect to H$_2$ of $7.9\times 10^{-5}$ in a shell spanning from 20 to 200\rstar{} with the parameters considered in the current work.
Since CN is a direct photodissociation product of HCN, this abundance can be taken as a lower limit for the HCN abundance, which is in good agreement with our results.
\citet{lindqvist_2000} found that the CN abundance distribution with respect to that of H$_2$ peaks at $5.5\times 10^{15}$~cm from the star, i.e., $\simeq 170\rstar$ (using the distance and stellar radius given in Table~\ref{tab:table1}).
They gave an HCN abundance greater than $6\times 10^{-5}$ also compatible with the abundance estimated in the current work.

\citet{schoier_2013} gave an abundance with respect to H$_2$ at the star of $3.0\times 10^{-5}$ and a photodissociation radius of $1.0\times 10^{16}$~cm ($\simeq 430\rstar$ with the stellar radius that we have adopted and the distance to the star that they chose, i.e., $\simeq 220$~pc).
The average abundance is thus $2.7\times 10^{-5}$.
Their maximum and average abundances are about a factor of 5 lower than ours, which make our abundance a priori incompatible with theirs.
However, the authors estimate the uncertainty of their abundance to be of a factor of 3.
This uncertainty along with the one derived for the outer CSE in the current work ($\simeq 50\%$; Table~\ref{tab:table2}), the different distances to the star adopted in both works, and the calibration errors related to the millimeter observations could make both results more compatible.

As we noted in Section~\ref{sec:molecular.emission}, the HCN column density in the CSE of \ycvn{} is surprisingly high compared to that in \irc{} \citep[$\simeq 2.1\times 10^{18}$ and $1.6\times 10^{19}$~cm$^{-3}$, respectively;][]{fonfria_2008}, considering the high disparity between their mass-loss rates \citep[$\simeq 1.4\times 10^{-7}$ and $2.7\times 10^{-5}$~\mlr;][]{guelin_2018}.
This fact is even clearer when their HCN abundances with respect to H$_2$ are compared: they are $\simeq 1.3\times 10^{-4}$ in the envelope of \ycvn{}  and $0.6-2.4\times 10^{-5}$ in \irc, if we correct the abundance derived by \citet{fonfria_2008} for a distance of 120~pc \citep{groenewegen_2012} and \citeauthor{guelin_2018}'s mass-loss rate.
Hence, the HCN abundance in the envelope of \ycvn{} is one order of magnitude higher than in \irc, the archetypal C-rich AGB (N-type) star.
Nevertheless, this can be explained if we consider that the HCN abundance grows from its photospheric value as the distance from the star increases while the CN abundance steeply decreases \citep{agundez_2020}.
The column density of photospheric CN in \ycvn{} is at least as high as in \irc{} (e.g., \citealt{barnbaum_1994,bakker_1997}; Fonfr\'ia et al., in prep.) but the H$_2$ column density in the envelopes of these stars are substantially different.

\subsubsection{The $^{12}$C/$^{13}$C isotopic ratio}
\label{sec:discussion.isotopic.ratio}

The fits to the HCN and H$^{13}$CN lines that we have done in this work confirms once more \citep{lambert_1986,olofsson_1993b,ohnaka_1999,schoier_2000,milam_2009,ramstedt_2014,abia_2017} that the isotopic ratio $^{12}$C/$^{13}$C in the envelope of \ycvn{} is typical of a J-type star.
Our result of 2.5 is in good agreement with the previous estimates, which range from 2 to 8.
Moreover, the two isotopic ratios that we have derived for the inner and outer envelope suggest that this quantity has not substantially changed over the last 350~yr.

\subsubsection{The extent of the HCN and abundance distribution}
\label{sec:discussion.extent}

The photodissociation radius obtained in the current work ranges from $\simeq 170$ to 380\rstar.
It is contained within the domain determined from prior observational estimates ($50-420\rstar$).
Yet, the data dispersion of the measures is too high and the agreement between these measures poor.

The CN($1-0$) interferometer observations analyzed by \citet{lindqvist_2000} indicate that the HCN abundance distribution, which has to be smaller than the CN abundance distribution, cannot extend further than $\simeq 430\rstar$ from \ycvn{} (using the distance to the star and stellar radius of Table~\ref{tab:table1}, and assuming a roughly spherically symmetric photodissociation shell).
In fact, the HCN abundance distribution is likely limited to within $200\rstar$ from the star, that is, inside the CN abundance peak.
However, the CN brightness distribution integrated over velocity is elongated along the north-south direction, and it consists of several bright spots at the systemic velocity, which suggests that the outer envelope of \ycvn{} can be irregular or clumpy.

In any case, our result is in better agreement with the largest photodissociation radius proposed by \citet{schoier_2013}, $\simeq 430\rstar$, and with the radius of the region that contains the CN emission \citep{lindqvist_2000} than with the rest of the estimates.
Nevertheless, the use of Eq.~7 of \citet{massalkhi_2018b} and the interstellar photodissociation rates for HCN in Table~B.1 of \citet{agundez_2018} give that the HCN abundance decreases by a factor of 10 with respect to its photospheric value at $\simeq 75\rstar$.
This radius is compatible with the shortest observational estimates.

This disagreement might indicate that either the interstellar UV radiation field is actually weaker around \ycvn's location in the Galaxy, or there is a dust shell optically thicker than expected that protects the HCN inside.
The latter scenario might be related to the increased density of dust grains that we propose to exist beyond $\simeq 200\rstar$ (Section~\ref{sec:continuum}).

\subsubsection{The kinetic, rotational, and vibrational temperatures}
\label{sec:discussion.temperatures}

In this work, we have adopted the kinetic temperature dependence over the outer layers of the envelope of \irc{} for the CSE of \ycvn{} based on their similar chemical composition and low gas density (Section~\ref{sec:envelope.model}).
Note however that the properties of the dust grains (e.g., size, expansion velocity) in both environments are different due to the particular physical and chemical conditions in their dust formation zones.
Hence, a detailed modeling of the thermal balance \citep[e.g.,][]{decin_2006}, which is out of the scope of the current work, may conclude that the kinetic temperature profile in \ycvn{} is different than assumed.

It can be argued that such a difference could significantly influence our results.
Nevertheless, the code that we have used relies on input excitation temperatures which have been implemented to be independent of the kinetic temperature (Appendix~\ref{sec:appendix.code}).
Consequently, \tk{} is only used as the excitation temperature by default and to calculate the thermal line width.
Our results indicate that the local line width is dominated by turbulence and the number of modeled lines is high enough to allow us to estimate the rotational and most of the vibrational temperatures.
Only a few \tvib{} are still considered as equal to \tk{} in the best fit model and all of them describe the population relation between adjacent vibrational states whose energy difference is only of a few tens of K (e.g., $2\nu_2(\sigma^+)$ and $2\nu_2(\delta)$; Table~\ref{tab:table2}), i.e., much smaller than the typical \tk{} in the region of the CSE where these vibrational states are populated.
Therefore, the effect of a change of \tk{} on our results is quite limited and can be neglected.

The rotational temperatures near the photosphere derived from the fits to the lines at 7.5 and $13-14~\mu$m differ by more than a factor of 2 between them.
In contrast, the \trot{} profiles beyond $\rdin\simeq 3.2\rstar$ are quite similar (see Table~\ref{tab:table2}) and can be assumed to be the same ($\trot\simeq 800 (\rdin/r)^{0.56}$~K).
The low rotational temperature in the acceleration zone ($r\simeq 1-3.2\rstar$) that affects the lines observed at $13-14~\mu$m implies that there is a substantial emission deficit in the high-$J$ lines formed in this shell.
The gas density in this region is $\sim 10^8-10^{10}$~cm$^{-3}$, high enough to roughly thermalize the rotational levels of HCN by collisions with H$_2$.
Hence, this low \trot{} may be explained by a clumpy or incomplete emitting shell in terms of the gas density or HCN abundance, an explanation already invoked by \citet{dinh-v-trung_2000} and compatible with our findings about the decrease of the HCN abundance close to the star.
In this scenario, the emission component of the high-$J$ lines would be more affected than those of the low-$J$ ones, which are essentially formed further from the star.

Most of the vibrational temperatures are $\lesssim 1500$~K, far below the effective stellar temperature ($\simeq 2900$~K).
However, the vibrational temperatures are biased by the bulk emission/absorption of the high excitation HCN lines located $\simeq 1\rstar$ beyond the photosphere.
The vibrational temperature across this shell can decrease very fast due to the typically high Einstein coefficients of the HCN ro-vibrational levels ($A_{ul}\simeq 0.5-5$~s$^{-1}$) making our result compatible with HCN under vibrational LTE at the photosphere.

Either way, our results indicate that the vibrational states are predominantly radiatively excited throughout most of the envelope in good agreement with the fact that the critical density for the ro-vibrational transitions \citep[with de-exciting collisional rates usually below $10^{-10}$~cm$^{3}$~s$^{-1}$; e.g.,][]{gonzalez-alfonso_2002,tobola_2008} is higher than at any region of the envelope ($n_\subscript{crit}\gtrsim 10^{10}$~cm$^{-3}$).
This critical density is typically of $10-100$ times higher than for the pure rotational lines, which explains the excitation differences between the rotational and vibrational temperatures.

Among the vibrational temperatures derived from our fitting process there is one clearly incompatible with the rest.
The \tvib[$2\nu_2(\delta)-2\nu_2(\sigma^+)$] as derived from the 7.5~$\mu$m lines is compatible with the kinetic temperature ($\simeq 2900-1300$~K for $r=1.0-3.2\rstar$), which is as expected for two vibrational states so close in energy ($\Delta E_\subscript{vib}\simeq 30$~K).
However, the \tvib{} derived from the $13-14~\mu$m is much lower ($\lesssim 50$~K).
Figure~\ref{fig:f6} shows how the model systematically overestimates the absorption of the lines of the Q branch of the $2\nu_2(\delta)-\nu_2(\pi)$ band by a factor of up to 2 while the lines of the R and P branches are satisfactorily well reproduced.
This is not physically possible as the lines of these three branches involve the same ro-vibrational levels.
Moreover, the code has been thoroughly tested over the years and the lines of the three branches have been calculated simultaneously with the same formulas.
This disagreement can be a consequence of inaccurate Einstein coefficients for the lines of the R or Q branches of the $2\nu_2(\delta)-\nu_2(\pi)$ band, as we also suggested during the analysis of the ro-vibrational diagram (Section~\ref{sec:molecular.emission}), but also the effect of radiative transfer mechanisms not considered in the envelope model, such as maser emission in high-$J$ lines or initially unexpected emission/absorption effects due to high line optical depths \citep[e.g.,][]{lacy_2013}.
Moreover, the ro-vibrational lines of the $4\nu_2(\sigma^+)-2\nu_2(\sigma^+)$ band might be affected by the Fermi resonance that exists between the vibrational states $\nu_1+\nu_2(\pi)$ and $4\nu_2(\sigma^+)$, which causes strong maser emission in the lines of the latter state and was detected in \ycvn{} by  \citet{schilke_2003}.

\subsubsection{Maser emission in the vibrational ground state and $\nu_2$}
\label{sec:masers}

\citet{dinh-v-trung_2000} did a deep analysis of the maser emission of the HCN line $v=0$ $J=1-0$ roughly explaining the overall line profile except for the central peak.
The pumping mechanism mostly relies on the vibrational excitation produced by photons involving the $\nu_1(\sigma^+)$\footnote{\citet{dinh-v-trung_2000} used a notation different than ours to name the vibrational modes of HCN. For them, the currently vibrational mode $\nu_1$ was $\nu_3$ and vice versa. We adopted the notation of, for instance, \citet{maki_2000}, which is the one also used in the HITRAN Database \citep{hitran}.} band at $3~\mu$m, where the stellar radiation field is strong (see Figure~\ref{fig:f4}).

Most of the bumps in the line profile were explained as blue- and red-shifted maser peaks related to each component of the hyperfine structure if the gas expansion velocity were 6~\kms.
This velocity is in good agreement with our conclusion that the maser emission is mostly produced at the end of the acceleration region.
The Doppler velocity of the red-shifted spike at +5~\kms{} with respect to the central peak of the line has not changed significantly in 20~yrs, which suggests that the gas acceleration mechanism has been stable during the last decades.
On the contrary, the intensities of several of these peaks do change over time, as it can be seen when the observation by \citeauthor{dinh-v-trung_2000} is compared with those by \citet{schoier_2013} and ours.
These variations may be related to the stellar pulsation period or they may indicate a change of the physical conditions in the maser pumping region due to the evolution of the ejected matter, which is fully accelerated in about 5~yr.
The rest of the HCN $v=0$ $J=1-0$ line shows broad components overlapped rather than spikes.
This is evidence of an extended maser emitting region comprising different expansion velocities, where the radiation is likely enhanced due to the high turbulent velocity prevailing in the acceleration zone (Section~\ref{sec:discussion.turbulence}).
In this scenario, the central peak is very likely produced at the stellar photosphere.

The HCN line $\nu_2^{1e}$ $J=3-2$ is undoubtedly a maser from an inspection of its profile: displaying two narrow peaks, and its intensity, which is almost as high as the line $J=3-2$ of the vibrational ground state.
Its comparison with the same line of H$^{13}$CN supports this conclusion as the HCN one is 150 times stronger even though it should be similar to the isotopic ratio $^{12}$C/$^{13}$C~$\simeq 2.5$, had both lines a thermal origin.
The HCN line $\nu_2^{1f}$ $J=3-2$ is about 15 times stronger than the corresponding H$^{13}$CN line, also stronger than expected. 
Therefore, the HCN $f$ line could a weak maser as well.

Contrary to what occurs in \irc, where the $\nu_2^{1e}$ $J=2-1$ maser is favored by the strong continuum emission at 7 and 14~$\mu$m \citep{lucas_1989,menten_2018}, the $\nu_2^{1e}$ $J=3-2$ maser in \ycvn{} is probably pumped by the continuum emission at 3~$\mu$m, as the continuum is significantly weaker at longer wavelengths (Figure~\ref{fig:f4}).
The hot band $\nu_1+\nu_2(\pi)-\nu_2(\pi)$, which is as strong as $\nu_1(\sigma^+)$, along with the fact that molecules in the ro-vibrational level $\nu_2^{1e}$ $J=1$ can be de-excited to the vibrational ground state, could result in population inversion.
This pumping mechanism is compatible with possible maser emission in the $\nu_2^{1e}$ $J=2-1$ line, which has not been observed in \ycvn.

\subsubsection{Reliability of the spatial asymmetries in the abundance distribution and gas expansion velocity field}
\label{sec:asymmetries}

The existence of asymmetries in the inner layers of the CSE comes from the incompatibilities found in the physical and chemical conditions derived from the mid-IR spectra of HCN.
The observations at 7.5 and 14.0~$\mu$m were acquired with the same instrument (SOFIA/EXES) under very similar conditions and along two consecutive nights.
The data at 13.1~$\mu$m were taken with IRTF/TEXES about two months prior.
In principle, it might be possible to find incompatibilities between the IRTF/TEXES data at 13.1~$\mu$m and the SOFIA/EXES data at 14.0~$\mu$m due to instrumental differences or effects related to the stellar pulsation phase but the results turned out to be highly compatible.
Nevertheless, the data sets acquired with SOFIA/EXES almost simultaneously are those which display the largest incompatibilities.
This suggests that the disagreements that we have found are very likely real.
In fact, \citet{ragland_2006} warned about the possible existence of asymmetries in the innermost envelope of this star.

At larger distances from the star, the velocity field of the expanding gas and the HCN abundance distribution might be asymmetric as well.
The lack of other purely thermal rotational lines in our study in addition to the HCN and H$^{13}$CN $v=0$ $J=3-2$ ones has brought some uncertainties in the determination of the physical conditions throughout the outer layers of the envelope and the photodissociation radius.
However, there is no reason to think that the red-shifted wing found in these lines is related to any instrumental issue or error during the observation process.
It can be argued that these wings are actually the hyperfine components with the lowest frequencies but, in this case, it is not possible to reproduce the observed profiles unless the red-shifted part of the lines are partially absorbed, something that would require an unlikely rotational temperature profile behind the star.

\section{Summary and conclusions}
\label{sec:conclusions}

We present new single-dish high spectral resolution observations of the AGB star \ycvn{} in the mid-IR ranges $7.35-7.60$, $13.00-13.25$, and $13.93-14.04~\mu$m acquired with SOFIA/EXES and IRTF/TEXES.
We have identified about 130 lines of HCN and H$^{13}$CN of the bands $\nu_2(\pi)$, $2\nu_2(\sigma^+)$, $2\nu_2(\sigma^+)-\nu_2(\pi)$, $2\nu_2(\delta)-\nu_2(\pi)$, $3\nu_2(\pi)-\nu_2(\pi)$, $3\nu_2(\pi)-2\nu_2(\sigma^+)$, $3\nu_2(\pi)-2\nu_2(\delta)$, $3\nu_2(\phi)-2\nu_2(\delta)$, $4\nu_2(\sigma^+)-2\nu_2(\sigma^+)$, and $4\nu_2(\delta)-2\nu_2(\delta)$, which display either P-Cygni or pure absorption profiles.
These data sets have been complemented with the pure rotational lines $J=1-0$ and $3-2$ in several vibrational states, which have been observed with the IRAM 30~m telescope at 1.12 and 3.38~mm, and the continuum emission taken with \textit{ISO}.

We have analyzed the HCN and H$^{13}$CN lines following two different approaches:
(1) an overall study based on a ro-vibrational diagram, and
(2) a detailed analysis in 1D and 2D that makes use of a code that reproduces the molecular and dust absorption and emission of the envelope of an AGB star \citep{fonfria_2008,fonfria_2014}.
The results derived from these analyses lead us to conclude that:
\begin{itemize}
\item The continuum emission can be described by a central star with an effective temperature of $\simeq 2900$~K and an optically thin dusty cocoon.
  The dust grains are mostly made of SiC in the inner layers of the envelope and of AC in the outer envelope.
The bulk of SiC condenses at $\simeq 3.0-3.5\rstar$ from the center of the star and the AC beyond $\simeq 200\rstar$.
At this distance, the density of dust grains seems to be about 15 times higher than predicted for a dusty isotropically expanding CSE.
\item The detailed modeling to the observed data and the ro-vibrational diagram suggest the existence of asymmetries in the physical and chemical conditions throughout the envelope.
\item The line profiles are compatible with a radial gas expansion velocity profile growing from $\simeq 0.5$ up to $\simeq 8$~\kms{} in the first 2\rstar{} beyond the photosphere.
  The terminal expansion velocity that we have used in our models ranges from 6 to 10~\kms{} along different directions but the actual maximum velocity might be even higher.
These variations might be related to higher velocity outflows.
\item The observed mid-IR lines are broader than expected.
  The best fits have been achieved with a line width at the stellar photosphere of $\simeq 10$~\kms, i.e., a turbulent velocity of $\simeq 6$~\kms.
Such a high turbulent velocity might be hiding other phenomena as high velocity matter ejections or photospheric movements related to the stellar pulsation or convection.
\item The HCN abundance with respect to H$_2$ is $\simeq 1.3\times 10^{-4}$ throughout most of the envelope.
The abundance may be significantly lower in the vicinity of \ycvn.
The average column density is $\simeq 2.1-3.5\times 10^{18}$~\cmm, only one order of magnitude lower than for \irc.
\item The rotational and vibrational temperatures of HCN are out of LTE throughout most of the envelope, which suggests that the collisions are too scarce to thermalize the gas even rotationally.
\item The pure rotational lines $v=0$ $J=1-0$ of HCN and H$^{13}$CN display maser emission \cite[e.g.,][]{dinh-v-trung_2000}.
  The $\nu_2^{1e}$ $J=3-2$ line has been found to show also evident maser emission.
  The $\nu_2^{1f}$ $J=3-2$ line might be a weak maser as well.
\item The isotopic ratio $^{12}$C/$^{13}$C is determined to be around 2.5 throughout the whole envelope in good agreement with previous measures.
\end{itemize}
All these conclusions lead us to think that \ycvn{} is a quite active star which could have developed structure in its envelope.
Due to the physical mechanisms that are producing the suggested asymmetries close to the star, the high temperatures, gas densities below $\sim 10^8-10^{10}$~cm$^{-3}$, the influence of the Galactic UV radiation field in the chemistry at different distances from the star, and its proximity, \ycvn{} is a good candidate to better understand the physical and chemical evolution of the envelopes of low mass-loss rate C-rich stars.

\begin{acknowledgements}

We thank the anonymous referee for their accurate and helpful comments provided during the reviewing of this manuscript.
The research leading to these results has received funding support from the European Research Council under the European Union's Seventh Framework Program (FP/2007-2013) / ERC Grant Agreement n. 610256 NANOCOSMOS.
EJM acknowledges financial support for this work through award \#06\_0144 which was issued by USRA and provided by NASA.
MJR and EXES observations are supported by NASA cooperative agreement 80NSSC19K1701.
MSG thanks Spanish MCIN through grant AYA2016-78994-P.
Based in part on observations made with the NASA/DLR Stratospheric Observatory for Infrared Astronomy (SOFIA).
SOFIA is jointly operated by the Universities Space Research Association, Inc. (USRA), under NASA contract NNA17BF53C, and the Deutsches SOFIA Institut (DSI) under DLR contract 50 OK 0901 to the University of Stuttgart.
Visiting Astronomer at the Infrared Telescope Facility, which is operated by the University of Hawaii under contract NNH14CK55B with the National Aeronautics and Space Administration.
This work is based on observations carried out under project numbers 155-16 and 048-17 with the IRAM 30~m telescope. IRAM is supported by INSU/CNRS (France), MPG (Germany) and IGN (Spain).
This publication makes use of data products from the Two Micron All Sky Survey, which is a joint project of the University of Massachusetts and the Infrared Processing and Analysis Center/California Institute of Technology, funded by the National Aeronautics and Space Administration and the National Science Foundation.
This publication makes use of data products from the Wide-field Infrared Survey Explorer, which is a joint project of the University of California, Los Angeles, and the Jet Propulsion Laboratory/California Institute of Technology, funded by the National Aeronautics and Space Administration. 

\end{acknowledgements}

\begin{appendix}

\section{Brief description of HCN}
\label{sec:intro.HCN}

Hydrogen cyanide is linear triatomic molecule that belongs to the C$_\infty$ group and displays a permanent dipole moment of 2.985~D \citep{ebenstein_1984}.
The vibration of its atoms can be described with the aid of four normal modes, two of which are degenerate and are combined into the bending mode $\nu_2(\pi)$.
These two degenerate normal modes involve atomic oscillations out of the molecular axis that allow the molecule to spin around it.
This results in the creation of a vibrational angular momentum that splits the rotational levels in sublevels with different $e-f$ parity \citep{brown_1975}.
The other two normal modes, $\nu_1(\sigma^+)$ and $\nu_3(\sigma^+)$, are stretching modes that can be essentially described as the HC--N and H--C vibrations along the molecular axis, respectively \citep[e.g.,][]{herzberg_1956}.

In the current work, we have adopted two different notations commonly used in Molecular Spectroscopy and Astrophysics to label a vibrational state: $v_1\nu_1+v_2\nu_2+v_3\nu_3(\ell^\pm)$ and $v_i\nu_i^{\ell p}$, where $i=1,2,3$.
The $v_i$ is the vibrational quantum number related to the $i$-th normal mode, $\ell$ the quantum number for the vibrational angular momentum, which can be expressed with the Greek letters $\sigma$, $\pi$, $\delta$, $\phi$, $\gamma$, \ldots{} for $\ell=0,1,2,3,4,\ldots$, ${}^\pm$ is the total parity of the molecule in this vibrational state (usually given only for the $\sigma$ states), and $p$ is the $e-f$ parity.
  The $v=0$ means the vibrational ground state (G.S.).
  The change of the rotational number in a transition is given by B$_p(J_\subscript{low})$ for the ro-vibrational lines, where B is the branch (P: $\Delta J=J_u-J_l=-1$, Q: $\Delta J=0$, and R: $\Delta J=+1$), and $J=J_u-J_l$ for the pure rotational ones.
More information about the selection rules and the line intensities of linear molecules can be found, e.g., in \citet{fonfria_2008}.

\begin{figure}
  \centering
  \includegraphics[width=0.475\textwidth]{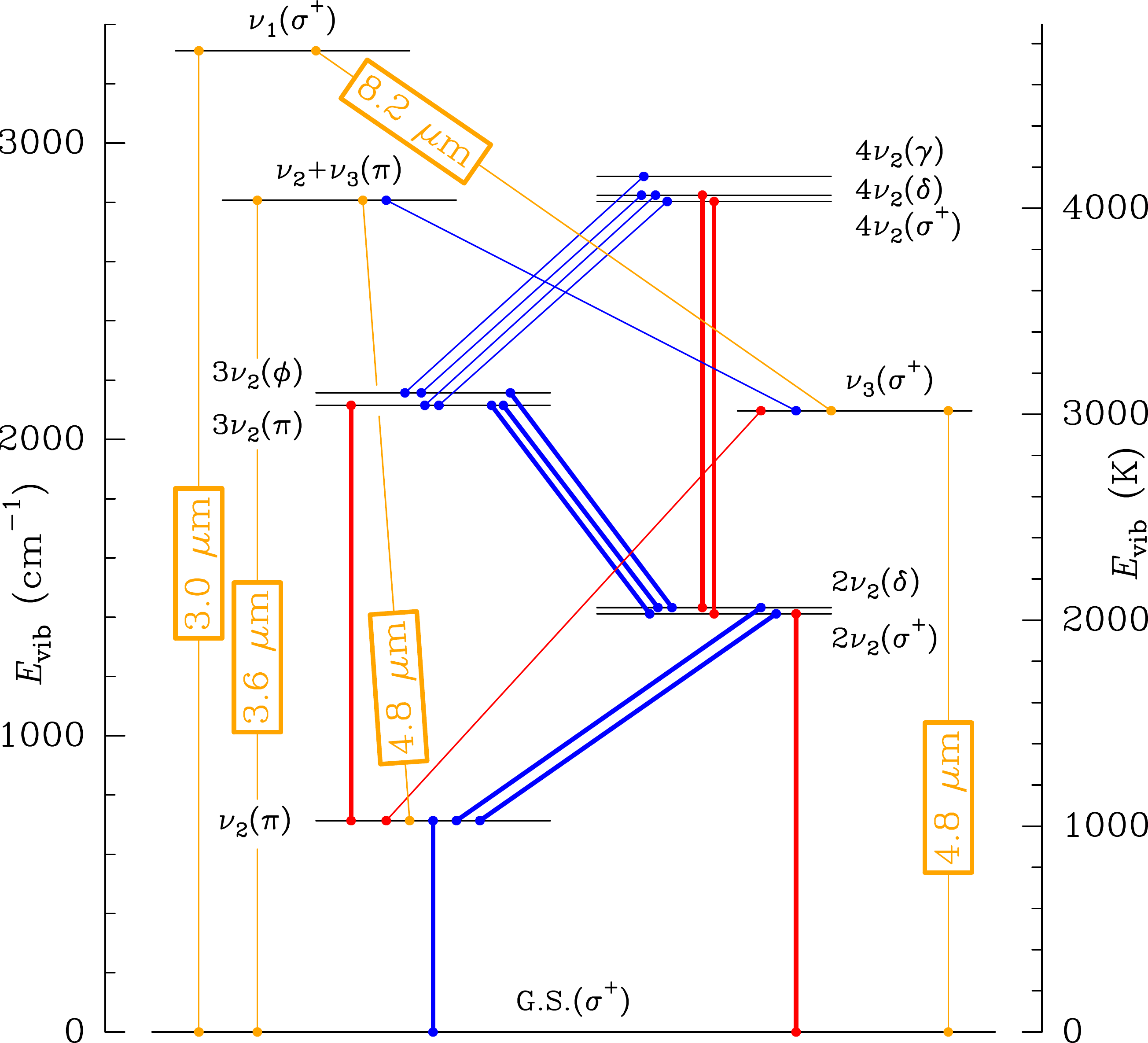}
  \caption{Vibrational energy diagram of HCN below 5000~K.
      The lines that connect the vibrational states represent the strongest allowed vibrational transitions.
      The transitions in blue lie around $13.5~\mu$m, in red around $7.5~\mu$m, and in orange at other wavelengths (3.0, 3.6, 4.8, and $8.2~\mu$m, depending on the transition).
      Those plotted with ticker lines have been detected in the spectra presented in the current work (Figures~\ref{fig:f1} and \ref{fig:f2}).
    }
  \label{fig:fa1}
\end{figure}

Figure~\ref{fig:fa1} contains a vibrational energy diagram for HCN to help understand the lines analyzed in the current work.

\section{Ro-vibrational diagrams}
\label{sec:rovib.diagram}

The absorption and emission components of a P-Cygni profile can be separated by assuming that they are produced in a layer with a constant temperature and density.
In this scenario, the solution of the radiation transfer equation is:
\begin{equation}
 F_\nu=I_{\nu,0}e^{-\tau_\nu}\Omega+S_\nu\left[1-e^{-\tau^\prime_\nu}\right]\Omega^\prime,
\end{equation}
where $F_\nu$ is the received flux, $I_{\nu,0}$ is the continuum emission, $\tau_\nu$ and $\tau^\prime_\nu$ are the optical depths related to the absorption and emission components, respectively, and $S_\nu$ is the source function that depends on the rotational and vibrational temperatures.
Each optical depth is affected by a Doppler shift.
The $\Omega=\Omega(\theta_s)=\frac{\pi}{4}(\theta_s^2+\theta_b^2)$ and $\Omega^\prime=\Omega(\theta_s^\prime)$ are the solid angles subtended by the absorbing and emitting regions (with sizes of $\theta_s$ and $\theta_s^\prime$, respectively), as seen by the telescope.
$\theta_b$ is the size of a point-like source, that can be calculated by summing in quadrature the size of the Point Spread Function (PSF) of the telescope, the atmospheric seeing, and the jittering of the telescope, if present.
Hence, for a normalized spectrum where the continuum flux is $F_{\nu,\subscript{cont}}=I_{\nu,0}\Omega$:
\begin{equation} F_\nu^\subscript{norm}=\frac{F_\nu}{F_{\nu,\subscript{cont}}}=e^{-\tau_\nu}+\frac{S_\nu}{I_{\nu,0}}\left[1-e^{-\tau^\prime_\nu}\right]\frac{\Omega^\prime}{\Omega}.
\end{equation}
To first approximation, the optical depths can be assumed to be Gaussian-like ($\simeq \tau_0 e^{-(\nu-\nu_0)^2/\sigma^2}/\sigma\sqrt{\pi}$).
The simplicity of the absorption component with respect to the emission one, more affected by unknowns, leads us to use the first one to derive reliable physical and chemical quantities from our observations.

We consider the opacity of a line \citep*[e.g.,][]{jacquemart_2001,simeckova_2006}:
\begin{equation}
  k_0=\frac{A_{ul}}{8\pi\nu^2 c}\frac{g_u}{Z}\frac{g_0}{g_l}\frac{n_l}{n_0}\left(1-\frac{g_l}{g_u}\frac{n_u}{n_l}\right),
\end{equation}
where $n_0$ is the numerical density of molecules in the lowest energy level, $n_l$, and $n_u$ the density of molecules in the lower and upper levels involved in the transition, and $Z$ is the total partition function.
The density fractions can be written by means of Boltzmann factors in this way:
\begin{equation}
  \frac{g_0}{g_l}\frac{n_l}{n_0}=e^{-hcE_{\subscript{vib,low}}/k_\subscript{B}T_{\subscript{vib},l}}\times e^{-hcE_{\subscript{rot,low}}/k_\subscript{B}T_{\subscript{rot},l}}
\end{equation}
and
\begin{equation}
  \frac{g_l}{g_u}\frac{n_u}{n_l}=e^{-hc\nu/k_\subscript{B}T_\subscript{exc}},
\end{equation}
where $E_{\subscript{vib,low}}$ and $E_{\subscript{rot,low}}$ are the vibrational an rotational energies of the lower ro-vibrational level involved in the transition ($E_{\subscript{low}}=E_{\subscript{vib,low}}+E_{\subscript{rot,low}}$), $T_{\subscript{vib},l}$ and $T_{\subscript{rot},l}$ ($T_\subscript{rot}$, hereafter) are the vibrational and rotational temperatures for the lower ro-vibrational level, respectively, and $T_\subscript{exc}$ is the excitation temperature of the line.

The integral of the optical depth over the frequency, $W$, derived from the absorption component of the observed line\footnote{Note that the integral of the optical depth over the frequency, $W=\int\tau_\nu d\nu$, can only be approximated by the integral over the absorption component of the line for optically thin lines. For optically thicker lines, the use of the integral over the absorption component of the line leads to wrong ro-vibrational diagrams due to the exponential character of the absorption.} can be related to these temperatures and the column density, $N_\subscript{col}$, through this formula:
\begin{align}
  \label{eq:rovib.diagram.step}
  \ln\left[\frac{W\nu^28\pi c}{A_{ul}g_u N_\subscript{col,0}}\right]= &
  \left\{
                                                                        \ln\left[\frac{N_\subscript{col}}{N_\subscript{col,0}Z}\left(1-e^{-hc\nu/k_\subscript{B}T_\subscript{exc}}\right)\right]\right. \nonumber \\
  & \left. -\frac{hcE_\subscript{vib,low}}{k_\subscript{B}T_{\subscript{vib},l}}
  \right\}
  -\frac{hcE_\subscript{rot,low}}{k_\subscript{B}T_\subscript{rot}}.
\end{align}
$N_{\subscript{col},0}$ is an arbitrary, constant column density introduced to produce dimensionless arguments for the logarithms.

The term $1-e^{-hc\nu/k_\subscript{B}T_\subscript{exc}}$ can be neglected for lines in the infrared if the excitation temperature $T_\subscript{exc}\ll hc\nu/k_\subscript{B}$, as it occurs in the outer shells of the envelopes of AGB stars \citep{fonfria_2017,fonfria_2018}.
However, in warmer shells its contribution can, in principle, be significant.
For our data, plotted in the ro-vibrational diagram that can be seen in Figure~\ref{fig:f5}, $1-e^{-hc\nu/k_\subscript{B}T_\subscript{exc}}$ varies less than 7\% due to its dependence on the frequency.
Thus, the effect of the frequency variation on the rotational temperature determination is by far much lower than the statistical noise derived form the linear fit to the scattered data sets in the ro-vibrational diagram and it can be ignored.
This term can therefore be evaluated at the average frequency of each set of lines that belong to a given band, $\bar\nu$.

The excitation temperature of the lines considered in this work can be expressed in terms of the vibrational and rotational temperature.
For a given line with a frequency $\nu$, considering the same rotational constant and rotational temperature for the vibrational states involved in the transition, and neglecting the contribution of the distortion and higher order rotational constants:
\begin{equation}
  \frac{hc\nu}{k_\subscript{B}T_\subscript{exc}}\simeq
  \frac{hc\Delta E_{\subscript{vib},ul}}{k_\subscript{B}T_{\subscript{vib},ul}}
 +\frac{hc\left\{B\left[J_u\left(J_u+1\right)-J_l\left(J_l+1\right)\right]\right\}}{k_\subscript{B}T_\subscript{rot}},
\end{equation}
where $\Delta E_{\subscript{vib},ul}$ and $T_{\subscript{vib},ul}$ are the energy difference and the vibrational temperature between the vibrational states involved in the transition.
For the lines of the $Q$ branch ($\Delta J=J_u-J_l=0$), $T_\subscript{exc}\simeq T_{\subscript{vib},ul}$.
However, for the lines of the $P$ and $R$ branches ($\Delta J=\pm 1$), $\left|B\left[J_u\left(J_u+1\right)-J_l\left(J_l+1\right)\right]\right|=2BJ_l$ and $2B\left(J_l+1\right)$, respectively.
On one hand, the rotational constant of HCN is $B\simeq 1.484775403$~\cm~$\simeq 2.2$~K \citep{maki_2000} and the rotational temperature in the shells where we expect the bulk absorption of HCN to be produced is of the order of $\simeq 1000$~K.
This means that $hc\left\{B\left[J_u\left(J_u+1\right)-J_l\left(J_l+1\right)\right]\right\}/k_\subscript{B}T_\subscript{rot}\ll 1$.
On the other hand, for the ro-vibrational lines of the HCN bands analyzed in this work (the fundamental band $\nu_2$ and the corresponding overtones and hot bands), the difference of vibrational energy between the upper and lower states is approximately a number of times the energy of the $\nu_2$ normal mode, which is $\simeq 713.4610$~\cm~$\simeq 1025$~K \citep{maki_2000}, and we expect the vibrational temperature to range from hundreds to a few thousands K.
Thus, we can say that $hc\Delta E_{\subscript{vib},ul}/k_\subscript{B}T_{\subscript{vib},ul}\gtrsim 1$ and $T_\subscript{exc}\simeq T_{\subscript{vib},ul}$ also for the lines of the $P$ and $R$ branches.
All these put together suggest that Eq.~\ref{eq:rovib.diagram.step} can be written as:
\begin{align}
  \label{eq:rovib.diagram.final}
  \ln\left[\frac{W\nu^28\pi c}{A_{ul}g_u N_\subscript{col,0}}\right]\simeq &
    \left\{
\ln\left[\frac{N_\subscript{col}}{N_\subscript{col,0}Z}\left(1-e^{-hc\bar\nu/k_\subscript{B}T_{\subscript{vib},ul}}\right)\right]\right. \nonumber \\
  & \left. -\frac{hcE_\subscript{vib,low}}{k_\subscript{B}T_{\subscript{vib},l}}
  \right\}
    -\frac{hcE_\subscript{rot,low}}{k_\subscript{B}T_\subscript{rot}}.
\end{align}
This formula can be used to do a linear fit to the data sets in the ro-vibrational diagram to derive the rotational temperature of each band.
The column density and the vibrational temperatures can be estimated from the $y$-intercepts derived from the fits to the data sets related to different bands.

\section{Further details of the radiative transfer code}
\label{sec:appendix.code}

The code allows us to determine the dust characteristics necessary to reproduce the low spectral resolution continuum emission.
Once the emission of the dusty component of the envelope is well reproduced, the high spectral resolution emission of the molecular gas is calculated.
This gas can involve one or several molecular species.
For each of them and in each position of the envelope, the code considers as many ro-vibrational levels as needed to properly describe the total population distribution.
The partition functions are calculated by direct summation over all the ro-vibrational levels until the increments per iteration are lower than a given value, typically $0.1\%$.
Due to the uncertainties of the ro-vibrational state-to-state collisional cross-sections and the high number of ro-vibrational levels usually involved in the calculations, the statistical equilibrium equations are not solved.
Instead of this, the populations of the ro-vibrational levels are controlled with input rotational and vibrational temperatures.
Ad-hoc excitation temperatures (positive and negative) can also be provided to modify the population ratio of certain individual levels allowing us, for instance, to invert their populations at will.
It is also possible to control the relative populations of the levels produced by the $\ell$-doubling mechanism that results in the $e$ and $f$ parities typical for linear molecules such as HCN \citep{brown_1975}.
These perturbed populations are also considered during the calculation of the partition function.
The kinetic temperature is not related to the excitation temperatures and it is only used to calculate the thermal broadening of the lines and the populations of the ro-vibrational levels under LTE.

The radiation transfer problem is solved by considering the effect of each gas and dust volume in the total emission for each impact parameter.
The code assumes that the dust grains are quiescent and the only interaction between them and the gas is radiative.
No heating-cooling balance is calculated and the mass-loss rate in form of dust cannot be estimated from our analysis.
Line blending is enabled by default.
The envelope emission is afterwards convolved with a Gaussian beam with the same half power beam width (HPBW) of the point spread function (PSF) of the telescope.
The resulting spectrum is convolved with another Gaussian if needed to mimic the resolving power provided by the instrument used during the observing run.
The calculation of the multidimensional synthetic emission is done by providing the 1D physical and chemical conditions in certain locations of the envelope.
The conditions in the rest of the CSE is derived by linear interpolation on the axial and polar angles.
Mapping the emission and various physical and chemical quantities is possible (Section~\ref{sec:asymmetric.model}).

\subsection{Molecular spectroscopic data and optical properties of dust grains}
\label{sec:spectroscopic.data}

The spectroscopic data used in this work for the line identification and modeling have been taken from three different sources.
Most of the HCN data have been retrieved from the HITRAN Database\footnote{\texttt{http://hitran.org/}} \citep{hitran}.
This data set has been complemented with additional data taken from the Exomol Database\footnote{\texttt{http://exomol.com/}} \citep{exomol}, useful to describe high excitation hot bands.
The spectroscopic data for the HCN rotational lines have been taken from the most recent version of the MADEX code (\citealt{cernicharo_2012}; Table~\ref{tab:table3}).
The optical properties of the solid state materials adopted to comprise the dust grains, i.e., amorphous carbon and amorphous SiC, have been taken from \citet{rouleau_1991} and \citet{mutschke_1999}, respectively.

\begin{table*}
\caption{Rest frequencies of pure rotational lines of HCN and H$^{13}$CN\label{tab:table3}}
\centering
\begin{tabular}{ccccccc}
\hline\hline
\multicolumn{3}{c}{Transition}  & \multicolumn{2}{c}{HCN} & \multicolumn{2}{c}{H$^{13}$CN}\\
mm Notation & mid-IR Notation & $F_\subscript{u}-F_\subscript{l}$ & No-HFS & HFS   & No-HFS & HFS \\
            &                 &                                 & (MHz)  & (MHz) & (MHz)  & (MHz)\\
\hline
$v=0$  $J=1-0$        & $\textnormal{G.S.}(\sigma^+)\textnormal{R}_e(0)$  &        & $\phantom{0}$88631.602  &                         & $\phantom{0}$86339.921  &     \\
                      &                                                   & $1-1$  &                         & $\phantom{0}$88630.415  &                         & $\phantom{0}$86338.735   \\
                      &                                                   & $2-1$  &                         & $\phantom{0}$88631.847  &                         & $\phantom{0}$86340.166   \\
                      &                                                   & $0-1$  &                         & $\phantom{0}$88633.936  &                         & $\phantom{0}$86342.255   \\
$v=0$ $J=3-2$         & $\textnormal{G.S.}(\sigma^+)\textnormal{R}_e(2)$  &        & 265886.433              &                         & 259011.798              &      \\
                      &                                                   & $3-3$  &                         & 265884.890              &                         & 259010.255   \\
                      &                                                   & $2-1$  &                         & 265886.188              &                         & 259011.552   \\
                      &                                                   & $3-2$  &                         & 265886.433              &                         & 259011.796   \\
                      &                                                   & $4-3$  &                         & 265886.499              &                         & 259011.862   \\
                      &                                                   & $2-3$  &                         & 265886.979              &                         & 259012.345   \\
                      &                                                   & $2-2$  &                         & 265888.522              &                         & 259013.886   \\
$\nu_2^{1e}$ $J=3-2$  & $\nu_2(\pi)\textnormal{R}_e(2)$                    &        & 265852.708              &                         & 258936.050              &     \\
$\nu_2^{1f}$ $J=3-2$  & $\nu_2(\pi)\textnormal{R}_f(2)$                    &        & 267199.282              &                         & 260224.813              &     \\
$2\nu_2^{0e}$ $J=3-2$ & $2\nu_2(\sigma^+)\textnormal{R}_e(2)$              &        & 267243.200              &                         & 260224.390              & \\
\hline
\end{tabular}
\tablefoot{The frequencies are included in the MADEX code \citep{cernicharo_2012} and have been derived from the fit to all the laboratory frequencies available hitherto in the literature and other data still unpublished involving many vibrational states (Cernicharo et al., in prep.).
The rest frequencies for the hyperfine components have been given for the $J=1-0$ and $3-2$ lines of the vibrational ground state.
The uncertainties are below 5~kHz for all the rest frequencies.
We have introduced a notation for the rotational transitions commonly used in works based on millimeter observations (column 1) and the transitions expressed with the notation that we are using in the current work for the mid-IR transitions (column 2).}
\end{table*}

\subsection{Parameter uncertainties}
\label{sec:uncertainties}

This code has also been used to estimate the uncertainties of the parameters.
Contrary to what happened with the line fitting process, the uncertainty estimate was done automatically.
If $\chi=\sqrt{\chi^2}$ and $\chi_\subscript{min}$ is the value for the minimum achieved for the best fit, $\chi-\chi_\subscript{min}$ assuming constant weights of $w_i=1/n$ for the $n$ frequency pixels is the mean difference between the synthetic spectrum and the best fit to the observations.
Hence, the $1\sigma$ uncertainty of a given parameter can be considered to be the maximum difference with respect to its best value (i.e., the value derived from the best fit) if $\chi-\chi_\subscript{min}$ is equal to the noise RMS of the observed spectrum.
Note that only synthetic lines are used in this process and neither the line density of the synthetic spectrum nor the high variety of intensities prevent the uncertainties to be achieved.

It is noteworthy that all the parameters are varied at the same time to uncover possible unknown dependencies but the parameter space is not fully explored.
  We follow the gradient of the parameter under study with respect to the rest of the parameters.
  This can be done as $\chi-\chi_\subscript{min}$ has been previously equaled to the noise RMS of the observations:
  \begin{equation}
    \chi(\mathbf{\nu},\mathbf{F_\subscript{norm}};\alpha_1,\ldots,\alpha_i,\ldots,\alpha_n)-\chi_\subscript{min}=\textnormal{RMS}
  \end{equation}
  where $\{\alpha_i\}_{i=1,n}$ are the free parameters of the model and $\mathbf{\nu}$ and $\mathbf{F_\subscript{norm}}$ are the frequency pixels and the observed normalized flux, respectively.
  Formally, the parameter $\alpha_i$ can be derived from the latter equation:
  \begin{equation}
\alpha_i=\chi^{-1}(\textnormal{RMS}+\chi_\subscript{min},\mathbf{\nu},\mathbf{F_\subscript{norm}};\alpha_1,\ldots,\alpha_{i-1},\alpha_{i+1},\ldots,\alpha_n).
  \end{equation}
However, the $\chi$ function cannot be analytically inverted and the gradient of each parameter is calculated numerically.
This method reveals the sensitivity of the envelope model to each parameter, which is reflected on its uncertainty.

\end{appendix}

\end{document}

%% file: definitions.tex
\newcommand{\subscript}[1]{\textnormal{\tiny{#1}}}
\newcommand{\rdout}{\ensuremath{R_\subscript{\textsc{III}}}}
\newcommand{\rdin}{\ensuremath{R_\subscript{\textsc{II}}}}
\newcommand{\rph}{\ensuremath{R_\subscript{ph}}}

\newcommand{\tk}{\ensuremath{T_\subscript{K}}}
\newcommand{\tvib}{\ensuremath{T_\subscript{vib}}}
\newcommand{\trot}{\ensuremath{T_\subscript{rot}}}
\newcommand{\rstar}{\ensuremath{R_\star}}
\newcommand{\kms}{km~s$^{-1}$}
\newcommand{\cm}{cm$^{-1}$}
\newcommand{\cmm}{cm$^{-2}$}

\newcommand{\mlr}{M$_\odot$~yr$^{-1}$}

\newcommand{\irc}{IRC+10216}
\newcommand{\rleo}{R~Leo}
\newcommand{\ycvn}{Y~CVn}

\newcommand{\natast}{Nature Astron.}

\newcommand{\jms}{J. Mol. Spect.}

\newcommand{\jpb}{J. Phys. B: At. Mol. Opt. Phys.}

\newcommand{\jai}{JAI}

\newcommand{\acet}{\ensuremath{\textnormal{C}_2\textnormal{H}_2}}